\documentclass[%
preprintnumbers,
nofootinbib,
 amsmath,amssymb,
 aps,
]{revtex4-2}

\usepackage{graphicx}
\usepackage{dcolumn}
\usepackage{bm}
\usepackage{comment}

\begin{document}

\preprint{KEK-TH-2527, J-PARC-TH-0288}

\title{Is the Aharonov-Bohm phase shift for a non-closed path a measurable quantity ?}

\author{Masashi Wakamatsu}
 \email{wakamatu@post.kek.jp}
\affiliation{%
KEK Theory Center, Institute of Particle and Nuclear Studies,
High Energy Accelerator Research Organization (KEK),
Oho 1-1, Tsukuba, 305-0801, Ibaraki, Japan
}%




\date{\today}

\begin{abstract}
There recently appear some interesting attempts to 
explain the AB-effect through the interaction between the
charged particle and the solenoid current mediated by the exchange of 
a virtual photon.
A vital assumption of this approach is that AB-phase shift is proportional to
the change of the interaction energy between the charged particle and
solenoid along the path of the moving charge.
Accordingly, they insist that the AB-phase change along 
a path does not depend on the gauge choice so that 
the AB-phase shift for a non-closed path is in principle measurable. 
We however notice the existence of two fairly different discussions on the
interaction energy between the solenoid and a charge particle, 
the one is due to Boyer and the other is due to Saldanha and others. 
In the present paper,  based on a self-contained quantum mechanical
treatment of the combined system of a solenoid, a charged particle, and 
the quantized electromagnetic fields, we show that
both interaction energies of Boyer and of Saldanha are in fact gauge invariant
at least for non-singular gauge transformations but they are destined to cancel 
each other. Our analysis rather shows that the origin of the 
AB-phase can be traced back to other part of our effective Hamiltonian.
Furthermore, based on the path-integral formalism with our effective Lagrangian, 
we explicitly demonstrate that the AB-phase shift for a non-closed path is not a 
gauge-variant quantity, which means that it would not correspond to direct
experimental observables.
\end{abstract}


\maketitle


\section{\label{Intro}Introduction}

Since it was first predicted theoretically \cite{ES1949,AB1959}, 
the Aharonov-Bohm effect
(AB-effect) has raised a lot of debates which concern theoretical
foundation of modern physics. (For review, see
\cite{Peshkin1981,OP1985,PT1989,WKZZ2018},
for example.)
A central question is whether the magnetic vector potential is a physical 
entity or it is just a convenient mathematical tool for calculating force 
fields \cite{Feynman1963,Konopinski1978,Semon-Taylor1996}.
Most popular interpretation of the AB-effect is that the vector potential
locally affects the complex phase of an electron wave function, thereby
causing a change of phase that can be measured through interference
experiments \cite{Tonomura1986,Osakabe1986}. 
Main objection to this understanding is based on the
fact that the vector potential is a gauge-dependent quantity 
so that  it has inherent arbitrariness.
By this reason, not a few researchers still hold on to nonlocal magnetic 
field interpretation \cite{Healey2022,ACR2016,Heras-Heras2022,Heras2022}. 
The ground of this field interpretation is that the AB-phase 
shift for a closed path can be expressed as a closed contour integration of 
the vector potential, so that, with the use of the Stokes theorem, it is 
eventually expressed with the magnetic flux penetrating the solenoid. 
Based on this fact, some authors insist that the AB-phase
shift is generated by the action of the magnetic field on the electron,
which are spatially separated from each other.  The great deficiency of such
nonlocality interpretation is that it appears to contradict the widely-accepted
locality principle of modern physics \footnote{Nowadays, it is widely
accepted that there is an unavoidable nonlocality in quantum mechanics
as exemplified by the famous Einstein-Podolsky-Rosen paradox \cite{EPR1935}
and Bell's inequality \cite{Bell1964}. 
However, it is also believed that this type of nonlocality has little to do with
the above-mentioned nonlocality in the interpretation of the AB effect}. 
On the other hand, despite the arbitrariness of the vector potential,
there exists some evidence that supports its physical reality as long as the 
theoretical interpretation of the AB-phase shift corresponding to a closed 
path \cite{WKZZ2018,Adachi1992,Stewart2003,Li2012}. 
It is based on the idea that the vector potential 
contains in itself a gauge-invariant piece which can never be eliminated 
by any {\it regular} gauge transformation.  
In fact, it has been argued by several authors that the AB-phase shift is
saturated by this gauge-invariant piece of the vector potential and
the remaining gauge arbitrariness has no influence on 
the AB-phase shift in such standard 
settings \cite{WKZZ2018,Stewart2003,Li2012}.

A new question arises, however, if one considers the AB-phase shift 
for a non-closed path.
This is because the contribution of the gauge-variant longitudinal component 
of the vector potential does not generally vanish for the AB-phase
shift for a non-closed path. In other words, the AB-phase shift 
corresponding to a non-closed path is usually believed to be 
a gauge-variant quantity. 
This indicates that it would not correspond to an observable quantity
as far as one believes the celebrated gauge principle.
It seems to us that this contradicts the recent claim by several researchers 
that the AB-phase shift for a non-closed path can in principle be 
observed \cite{Boyer1971,Vaidman2012,Kang2017}
\nocite{Marletto-Vedral2020,Saldanha2021,LHK2022}. 
We recall that a vital assumption in their argument is that the 
AB-phase shift is proportional to the change of 
interaction energy between a charged particle and the current of the 
solenoid mediated along the path of a moving change.
According to them, since the change of energy along such a path is 
a gauge-invariant quantity, the AB-phase shift along such non-closed 
path is also gauge invariant, so that it can in principle be observed, . 
This is an astonishing conclusion, because, if it were true, it would mean that
the AB-effect is not necessarily a {\it topological} phenomena as is
widely advertised \cite{Kobe1991,CLBNGK2019}. 
In view of the impact of the conflict above,
we think it very important to elucidate how to dissolve this contradiction. 
To get a clear answer to this question is the main purpose 
of the present paper. 

\vspace{2mm}
The paper is organized as follows.
First, in sect.\ref{sect2}, we briefly review some recent papers which
claim the observability of the AB-phase shift for a non-closed path.
In sect.\ref{sect3}, to inspect the validity of such claims, 
we build up a self-contained quantum mechanical treatment 
of the combined system of a solenoid and a charged particle
coupled to the quantized electromagnetic fields.. 
The analysis will be done based on two forms of physically equivalent 
Hamiltonians constructed there. This quantum mechanical analysis
is compared with the analyses by Boyer \cite{Boyer1971}, 
Saldanha \cite{Saldanha2021} and others \cite{Kang2017,Marletto-Vedral2020}, 
who claim that the AB-phase shit for a 
non-closed path is gauge invariant and it can in principle be 
observed. In sect.\ref{sect4}, using the path integral formalism based on
one form of our Lagrangians, we try to show that 
the AB-phase for a non-closed path is certainly a gauge-dependent 
quantity at variance with the recent claims mentioned above.
Sect.\ref{sect5} summarizes what we can conclude from the present 
investigation.

\section{\label{sect2}Brief review of recent proposals for measuring the AB-phase
shift for a non-closed path}

As discussed in Introduction, a widely-believed understanding
is that the AB-phase shift corresponding to a non-closed path
is a gauge-variant quantity, so that it
would not correspond to an observable quantity.
Recently, however, several authors claim that it can in principle
be observed, and have proposed several specific measurement methods
\cite{Kang2017,Marletto-Vedral2020,Saldanha2021}.
For example, Marletto and Vedral suggest that, to experimentally access 
the AB-phase shift corresponding to a {\it non-closed path}, one should consider 
a superposition of two different sharp position states and tell the AB-phase 
from the phase difference in this superposition \cite{Marletto-Vedral2020}. 
They further insist that this phase difference can in principle
be reconstructed by another reference electron in addition to a primary
electron and local tomography. 
In this scenario, however, Marletto and Vedral implicitly assume
that the AB-phase shift for a non-closed path is {\it gauge-invariant},
which seems to be inspired by Boyer's old work \cite{Boyer1971}.
Boyer's original treatment is based on the framework of classical
electrodynamics. 
(In the present paper, to avoid unnecessary notational complexity, we use
natural units, i.e. the Heaviside-Lorenz units with $c = 1$.)
He assumes that the AB-phase shift for some
non-closed path is proportional to the change of the interaction
energy between the magnetic field $\bm{B}^s$ generated by the
current of an infinitely-long solenoid and the magnetic field $\bm{B}^\prime$
generated by a moving charge with a constant velocity $\bm{v}$ as
\begin{equation}
 \Delta \phi_{AB} \ \propto \ \Delta \varepsilon \ = \ 
 \int d^3 x \,\,\bm{B}^s (\bm{x}) \cdot 
 \bm{B}^\prime (\bm{x}, t) .
\end{equation}
Here, $\bm{B}^s (\bm{x})$ generated by the surface current 
$\bm{j} (\bm{x})$ of the solenoid satisfies the equation
\begin{equation}
 \nabla \times \bm{B}^s (\bm{x}) \ = \ \bm{j} (\bm{x}) .
\end{equation}

\vspace{2mm}
\noindent
On the other hand, in the nonrelativistic limit, the electric and magnetic fields
generated by a moving charge with a {\it constant velocity} $\bm{v}$ is known
to be given as \cite{Heaviside1888,Jefimenko1994,RPS2006}
\begin{eqnarray}
 \bm{E}^\prime (\bm{x}, t) &=& \frac{e \,(\bm{x} - \bm{x}^\prime)}
 {\vert \bm{x} - \bm{x}^\prime \vert} \ = \ - \,\nabla \,
 \frac{e}{\vert \bm{x} - \bm{x}^\prime \vert} , \\
 \bm{B}^\prime (\bm{x}, t) &=& \bm{v} \times \bm{E}^\prime (\bm{x}, t) ,
\end{eqnarray}
with $d \bm{x}^\prime (t) / d t = \bm{v}$. (We recall that Boyer started 
with the relativistic expression for the electric and magnetic field, but he 
eventually took non-relativistic limit \cite{Boyer1971}.)
The interaction energy is then rewritten as \footnote{The deformation here
is possible, under the nonrelativistic treatment of the charged particle moving 
with a constant velocity. In fact, in the general case, the velocity vector depends
on the retarded time $v = v (t_r)$ and extracting it from volume integrations
is not possible, as it depends on the field point $\bm{x}$ through 
$t_r = t - \vert \bm{x} - \bm{x}^\prime \vert / c$.}
\begin{equation}
 \Delta \varepsilon \ = \ \frac{1}{4 \,\pi} \,\int \,d^3 x \,\,
 \bm{B}^s \cdot
 ( \bm{v} \times \bm{E}^\prime ) 
 \ = \ \bm{v} \cdot \frac{1}{4 \,\pi} \,\int \,d^3 x \,\,
 \bm{E}^\prime \times \bm{B}^s .
\end{equation}
Here, the relevant integral can be transformed as
\begin{eqnarray}
 &\,& \frac{1}{4 \,\pi} \,\int d^3 x \,\,
 \bm{E}^\prime (\bm{x}) \times \bm{B}^s (\bm{x})
 \, = \, - \,\frac{1}{4 \,\pi} \,\int d^3 x \,\,
 \left( \nabla \frac{e}{\vert \bm{x} - 
 \bm{x}^\prime \vert} \right) \times \bm{B}^s (\bm{x}) \nonumber \\
 &=& - \,\frac{e}{4 \,\pi} \int_S d S \,\left( 
 \frac{1}{\vert \bm{x} - \bm{x}^\prime \vert} \,\,\bm{n} \times
 \bm{B}^s (\bm{x}) \right)
 + \frac{e}{4 \,\pi} \int d^3 x \,
 \frac{1}{\vert \bm{x} - \bm{x}^\prime \vert} \,
 \nabla \times \bm{B}^s (\bm{x}). \hspace{8mm}
\end{eqnarray}
The surface term above can safely be dropped, since the magnetic field
by the solenoid current is confined inside the solenoid.
Then, using $\nabla \times \bm{B}^s (\bm{x}) = \bm{j} (\bm{x})$, he
obtains
\begin{equation}
 \frac{1}{4 \,\pi} \,\int \,d^3 x \,\,
 \bm{E}^\prime (\bm{x}) \times \bm{B}^s (\bm{x})
 \ = \ \frac{e}{4 \,\pi} \,\int \,d^3 x \,\,
 \frac{\bm{j} (\bm{x})}{\vert \bm{x} - \bm{x}^\prime \vert} .
\end{equation}
In this way, Boyer arrives at a remarkable 
relation \footnote{
There is some delicacy concerning the convergence of the integral
given by Eq.(\ref{current_integral}). Strictly speaking, this integral diverges
for an infinitely-long solenoid. On the other hand, for a very long but
finite-length solenoid as assumed by Boyer, the magnetic field outside
the solenoid does not exactly vanish and the form of $\bm{A}^{(S)}$
appearing in Boyer's paper is justified only approximately.
Since the discussion of this problem is a little involved, it is
forwarded to an Appendix (\ref{AppA}).}
\begin{equation}
 \Delta \varepsilon (\mbox{\tt Boyer}) \ = \ e \,\bm{v} \cdot 
 \bm{A}^{(S)} (\bm{x}^\prime) ,  \label{energy_Boyer}
\end{equation}
with
\begin{equation}
 \bm{A}^{(S)} (\bm{x}^\prime) \ = \ \frac{1}{4 \,\pi} \,\int \,d^3 x \,\,
 \frac{\bm{j} (\bm{x})}{\vert \bm{x}^\prime - \bm{x} \vert}.
 \label{current_integral}
\end{equation}

\vspace{2mm}
\noindent
Furthermore, just below Eq.(28) in his paper, Boyer stated 
as 

\begin{itemize}
\item[(1)] The $\bm{A}^{(S)}$ appearing in his interaction energy may 
be recognized as just the magnetic vector potential in the {\it Coulomb gauge}. 
\end{itemize}

\vspace{1mm}
\noindent
On the other hand, he seems to take for granted that

\begin{itemize}
\item[(2)] The interaction energy $\Delta \varepsilon ({\tt Boyer})$ is a
{\it gauge-choice independent} quantity.
\end{itemize}

\vspace{2mm}
\noindent
Slightly worrisome here is the mutual consistency of these two
observations.  Certainly, the expression for $\bm{A}^{(S)} (\bm{x})$
given by  Eq.(29) in his paper looks seemingly gauge invariant, since it is
expressed with the convolution integral of the {\it gauge-invariant} solenoid
current.  On the other hand, the startement (1) might give an impression that
the expression (29) for Boyer's energy is derived basically within the framework of
Coulomb gauge.
If it takes a different form in other choice of gauge, we cannot say that it is 
gauge independent. 
Since this is a little delicate problem, which deviates from the main 
discussion in this section, it is explained in Appendix \ref{AppB}.

\vspace{2mm}
\noindent
Returning to the main issue,
Boyer made step further by assuming that the AB-phase shift $\phi_{AB}$
is {\it proportional} to the change of the above interaction energy along the path of 
the moving charge, even when the path is {\it not} necessarily {\it closed}.
According to him, since this energy change is a gauge-invariant quantity,
the AB-phase shift for such a non-closed path is independent of
the gauge choice, so that the corresponding AB-phase can in principle be observed.
Undoubtedly, this viewpoint is also shared in the paper by Marletto and 
Vedral \cite{Marletto-Vedral2020}. We shall discuss later that
this claim is not justified.

\vspace{2mm}
\noindent
Partially motivated by the above-mentioned observation by Boyer,
several researchers investigated the interaction energy between the solenoid 
and a moving charge within the framework of quantum 
electrodynamics \cite{Saldanha2021,Santos-Gonzalo1999,LHK2022}.
Most comprehensive for our later discussion is the analysis by 
Saldanha \cite{Saldanha2021}. 
(We point out that his quantum mechanical analysis was done on the basis of 
the preceding work by Santos and Gonzalo \cite{Santos-Gonzalo1999}.)
Saldanha evaluated the interaction energy between the solenoid and the
charged particle mediated by the exchange of a virtual photon
within the framework of the quantum electrodynamics in the {\it Lorenz gauge}.
He starts with the {\it local} interaction between the quantized electromagnetic
field $\bm{A}$ and the solenoid current $\bm{j}$ given as
\begin{equation}
 V_1 \ = \ - \,\int \,d^3 x \,\,\bm{j} (\bm{x}) 
 \cdot \bm{A} (\bm{x}) ,
\end{equation}
where
\begin{equation}
 \bm{A} (\bm{x}) \ = \ \int \,d^3 k \, \,
 \sqrt{\frac{1}{(2 \,\pi)^3 \,2 \,\omega}} \,\,\,
 \sum_{\lambda = 0}^3 \,\,
 \bm{\epsilon}_{\bm{k},\lambda} \,\,
 a (\bm{k},\lambda) \,\,e^{\,i \,\bm{k} \cdot \bm{x}} \ + \ h.c.,
\end{equation}
with $\omega = \vert \bm{k} \vert$. Here, $\bm{\epsilon}_{\bm{k},\lambda}$
and $a (\bm{k},\lambda)$ respectively stand for the polarization
vector and annihilation operator of a photon with momentum
$\bm{k}$ and polarization $\lambda$. The polarization indices
$\lambda = 1,2$ correspond to the transverse photon,
$\lambda = 3$ to the longitudinal photon, and $\lambda = 0$ to
the scalar photon, respectively.
On the other hand, the {\it local} interaction between the electromagnetic 
field and a moving charge is given by
\begin{equation}
 V_2 \ = \ - \,e \,\bm{v} \cdot \bm{A} (\bm{x}^\prime) ,
\end{equation}
where $\bm{v}$ is the velocity of the moving charge.
Next, based on the framework of 2nd order perturbation theory, the 
interaction energy is evaluated as
\begin{equation}
 \Delta \varepsilon \ = \ \sum_{n \neq 0} \,
 \frac{\vert \langle n \,\vert \,V_1 \ + \ V_2 \,\vert \,vac \rangle \vert^2}
 {E^{(0)}_0 \ - \ E^{(0)}_n} ,
\end{equation}
where $\vert vac \rangle \equiv \vert n = 0 \rangle$ represents the 
vacuum of the photon, while
$\vert n \rangle$ with $n \neq 0$ stand for the exited states
of the free electromagnetic Hamiltonian.
Retaining only the one-photon intermediate states for $\vert n \rangle$
and dropping the self-energy terms, $\Delta \varepsilon$ reduces to
\begin{equation}
 \Delta \varepsilon \ = \ \mbox{Re} \,\left[ \,\int \,d^3 k \,
 \sum_\lambda \,
 \frac{\langle vac \,\vert \,V_2 \,\vert \,\bm{k}, \lambda \rangle \,
 \langle \bm{k}, \lambda \,\vert \,V_1 \,\vert vac \rangle}
 {- \,\omega} \,\right] \ + \ \left( V_1 \leftrightarrow V_2 \right) ,
\end{equation}
where $\vert \bm{k}, \lambda \rangle$ represents a one-photon state 
with the momentum $\bm{k}$ and the polarization $\lambda$.
Using
\begin{eqnarray}
 \langle \bm{k}, \lambda \,\vert \,V_1 \,\vert vac \rangle &=& 
 - \, \int d^3 x \,\,\,\int \,d^3 k \,
 \sqrt{\frac{1}{(2 \,\pi)^3 \,2 \,\omega}} \,\,\,\bm{j} (\bm{x}) \cdot 
 \bm{\epsilon}_{\bm{k}, \lambda} \,\,e^{\,- \,i \,\bm{k} \cdot \bm{x}} , \\
 \langle vac \,\vert \,V_2 \,\vert \,\bm{k}, \lambda \rangle &=&
 - \,e \,\,\sqrt{\frac{1}{(2 \,\pi)^3 \,2 \,\omega}} \,\,\,
 \bm{v} \cdot \bm{\epsilon}_{\bm{k}, \lambda} \,\,
 e^{\,i \,\bm{k} \cdot \bm{x}^\prime}.
\end{eqnarray}
and carrying out the polarization sum for the virtual photon states
with
\begin{equation}
 \sum_{\lambda = 0,1,2,3} \,\epsilon^\mu_{\bm{k}, \lambda} \,
 \epsilon^\nu_{\bm{k}, \lambda} \ = \ - \,g^{\mu \nu} ,
\end{equation}
Saldanha obtains
\begin{equation}
 \Delta \varepsilon \ = \ - \,e \,\,\mbox{Re} \,\left\{ \,\int \,d^3 x^\prime \,
 \left[ \int \,d^3 k \,\,\frac{e^{\,i \,\bm{k} \cdot (\bm{x}^\prime - \bm{x})}}
 {(2 \,\pi)^3 \,k^2} \,\right] \,\bm{v} \cdot \bm{j} (\bm{x}) \right\} . 
\end{equation}
%
Then, with the use of the identity
\begin{equation}
 \int \,\frac{d^3 k}{(2 \,\pi)^3} \,\,
 \frac{e^{\,i \,\bm{k} \cdot (\bm{x}^\prime - \bm{x})}}
 {k^2} \ = \ \frac{1}{4 \,\pi} \,\,
 \frac{1}{\vert \bm{x}^\prime - \bm{x} \vert} ,
\end{equation}
he eventually arrives at
\begin{equation}
 \Delta \varepsilon \ = \ - \,e \, \bm{v} \cdot 
 \bm{{\cal A}} (\bm{x}^\prime) ,
\end{equation}
with
\begin{equation}
 \bm{{\cal A}} (\bm{x}^\prime) \ = \ 
 \frac{1}{4 \,\pi} \,\int \,d^3 x^\prime \,\,
 \frac{\bm{j} (\bm{x})}{\vert \bm{x}^\prime - \bm{x} \vert} .
\end{equation}
With the standard setting of an extremely long solenoid, the vector
potential resulting from the above integral is nothing but
the axially-symmetric potential, i.e. $\bm{{\cal A}} (\bm{x}^\prime) =
\bm{A}^{(S)} (\bm{x}^\prime)$, which was also the case in Boyer's
analysis \cite{Boyer1971}.
We point out that, although the above result by Saldanha was obtained 
in the Lorenz gauge \cite{Saldanha2021}, it can be easily verified that 
exactly the same answer is obtained also in the Coulomb gauge.
The key ingredient here is the conservation law 
$\nabla \cdot \bm{j} (\bm{x}) = 0$
for the steady current of the solenoid.
At any rate, the above interaction energy obtained by Saldanha also
looks gauge-invariant. Based on this observation, Saldanha also proposed
a specific measurement method of the AB-phase shift corresponding
to a non-closed path. (See Fig.2 and the explanation in the caption
in his paper \cite{Saldanha2021}.) Besides, he also gave a theoretical 
prediction, which he would expect in such measurement, given as
\begin{equation}
 \Delta \phi_{AB} \ = \ \vert e \vert \,\int_{\tt path \ a} \,
 \bm{A}^{(S)} (\bm{x}^\prime) \cdot d \bm{x}^\prime \ - \ 
 \vert e \vert \,\int_{\tt path \ b} \,
 \bm{A}^{(S)} (\bm{x}^\prime) \cdot d \bm{x}^\prime \ = \ 
 \theta \,\,\frac{\vert e \vert \,\Phi}{2 \,\pi}.
\end{equation}
For the meaning of the path a and path b as well as
the meaning of the angle $\theta$, see Fig.2 in \cite{Saldanha2021}.
Somewhat strangely, if this prediction of Saldanha were confirmed
experimentally, it seems to us that it amounts to observing a
particular form of vector potential configuration in sharp contrast
to the widely accepted perception that the vector potential
does not correspond to an observable.

\vspace{2mm}
At this point, it is interesting to compare the interaction energy of
Boyer and that of Saldanha. They are given by
\begin{eqnarray}
 \Delta \varepsilon ({\tt Boyer}) &=& + \,e \,\bm{v} \cdot
 \bm{A}^{(S)} (\bm{x}^\prime), \\
 \Delta \varepsilon ({\tt Saldanha}) &=& - \,e \,\bm{v} \cdot
 \bm{A}^{(S)} (\bm{x}^\prime),
\end{eqnarray}
with
\begin{equation}
 \bm{A}^{(S)} (\bm{x}^\prime) \ = \ 
 \frac{1}{4 \,\pi} \,\int \,d^3 x \,\,
 \frac{\bm{j} (\bm{x})}{\vert \bm{x}^\prime - \bm{x} \vert} .
\end{equation}
As already pointed out, Boyer's analysis based on the
classical electromagnetism and Saldanha's analysis based on
the quantum electrodynamics give almost the same form of answer for
the interaction energy between the solenoid current and a moving charge.
Curiously, however, there is a delicate difference between their
results. That is, the overall {\it sign} of the interaction energy is just opposite,
which also means that the predictions for the AB-phase shift would
be different in sign.
To our knowledge, no one has paid attention to this  subtle but remarkable
difference. 
We are thus naturally led to the following questions.

\begin{enumerate}
\item Is there any meaning in the strange sign difference between
the interaction energies of Boyer and of Saldanha ?
\vspace{2mm}
\item Is the assumption that the AB-phase shit is proportional to
the interaction energy between the solenoid and a moving charge
justified, to begin with ?
\vspace{2mm}
\item Is the AB-phase shift for a non-colsed path a measurable
quantity, at all ? 
\end{enumerate}

\vspace{0mm}
\noindent
We try to answer these questions in the next two sections.

\section{\label{sect3}Self-contained quantum mechanical treatment of the combined
system of the solenoid, the moving charge, and the electromagnetic 
fields}

In pursuit of the root of the mysterious observation made at the end of the 
last section, here we propose to work in a self-contained treatment 
of the combined system of the moving charge and the solenoid, and
the electromagnetic fields.
The basic assumptions in our treatment are the nonrelativistic treatment
of charged particle motion, the 1st-order perturbation theory 
in the small charge of the moving particle, free field quantization of the
electromagnetic fields, neglect of self-interactions, etc.

\vspace{2mm}
\noindent
The mass, the charge and the coordinate of a moving particle are respectively
denoted as $m$, $e$, and $\bm{q}$. This charged particle not only interacts with
the electromagnetic fields, but also subjects to the action of a certain potential 
$V (\bm{q})$, which is supposed to work for preventing the charged
particle from entering the interior of the solenoid. The electromagnetic field
also couples to a given external current $j^\mu_{ext} (x) = 
(0, \bm{j}_{ext} (\bm{x}))$, where
$\bm{j}_{ext} (\bm{x})$ stands for the uniform {\it steady} current flowing 
on the surface of an extremely long straight solenoid oriented along the
$z$-axis.
The equations of motion for the charged particle and the electromagnetic
fields can straightforwardly be written down as follows :
\begin{subequations} \label{allA}
 \begin{align}
 &\, m \,\frac{d^2 \bm{q}}{d t^2} \ = \ - \,\frac{\partial V  (\bm{q})}{\partial \bm{q}}
 \ + \ 
 e \,\left( \bm{E} (\bm{q}, t) \ + \ \dot{\bm{q}} \times \bm{B} (\bm{q}, t) \right), 
 \label{first equation A} \\
 &\, \nabla \cdot \bm{E} (\bm{x}, t) \ = \ \tilde{j}^0 (\bm{x}, t) , 
 \label{second equation A} \\
 &\, \nabla \times \bm{B} (\bm{x}, t) \ = \ 
 \tilde{\bm{j}} (\bm{x}, t) \ + \ \bm{j}_{ext} (\bm{x}) \ + \ 
 \frac{\partial \bm{E} (\bm{x}, t)}{\partial t} \label{third equation A},
 \end{align}
\end{subequations}
with $\dot{\bm{q}} \equiv d \bm{q} / \partial t$.
Here, $\tilde{j}^\mu (x) = (\tilde{j}^0 (x), \tilde{\bm{j}} (x))$ represents the 
four-current density generated by the moving charge with a constant velocity 
$\dot{\bm{q}} = \bm{v}$ given as
\begin{eqnarray}
 \tilde{j}^0 (\bm{x}, t) &=& e \,\delta (\bm{x} - \bm{q}) , \\
 \tilde{\bm{j}} (\bm{x}, t) &=& e \,\dot{\bm{q}} \,\delta (\bm{x} - \bm{q}) .
\end{eqnarray}
It is a textbook exercise to show that the above equations of motion 
can be derived from the following Lagrangian $L_A$ : 
\begin{eqnarray}
 L_A &=& \frac{1}{2} \,m \,\dot{\bm{q}}^2 \ - \ V (\bm{q}) \ - \ 
 \frac{1}{4} \,\int \,d^3 x \,F_{\mu \nu} \,F^{\mu \nu} 
 \ - \ 
 \int \,d^3 x \,\left( \tilde{j}_\mu (x) + j^{ext}_\mu (x) \right) \,A^\mu (x) ,
\end{eqnarray}
where $F_{\mu \nu} (x) = \partial_\mu \,A_\nu \, - \, \partial_\nu \,A_\mu$
is the familiar field-strength tensor of the electromagnetic field.
To obtain the Hamiltonian corresponding to the above Lagrangian $L_A$
by means of the standard canonical procedure, we need to fix gauge.
For simplicity, here we choose the Coulomb gauge specified by 
$\nabla \cdot \bm{A} = 0$. 
This leads to the following Hamiltonian $H_A$ :
%
%
%
\begin{eqnarray}
 H_A &=& \frac{1}{2 \,m} \,\left( \,\bm{p} - e \,\bm{A} (\bm{q}, t) \right)^2
 \ + \ V (\bm{q})  \ + \ \int \,d^3 x \,\,\frac{1}{2} \, 
 \left( \bm{E}_\perp^2 \ + \ \bm{B}^2 \right) 
 \ - \ e \,\tilde{A}^0 (\bm{q}, t)  \ - \ 
 \int \,d^3 x \,\,\bm{j}_{ext} \cdot 
 \bm{A} (\bm{x}, t) . 
 \ \ \ \ \ \label{HA}
\end{eqnarray}
Here, $\tilde{A}^0 (\bm{q}, t)$ is the Coulomb potential generated by
the charge density $\tilde{j}^0 (\bm{x}, t)$ of the moving charge.
$\bm{E}_\perp$ represents the transverse component of the
electric field $\bm{E}$, 
while the longitudinal component of $\bm{E}$ is given by
\begin{equation}
 \bm{E}_\parallel (\bm{x}, t) \ = \ - \,\nabla \tilde{A}^0 (\bm{x}, t).
\end{equation}
Since the term $- \,e \,\tilde{A}^0 (\bm{q}, t)$ in the above $H_A$ plays no 
essential role in the following discussion, we simply drop it below.
(To be more precise, $\bm{E}_\perp + \bm{E}_\parallel$ represents
the electric field $\bm{E} (\bm{q}, t)$ appearing in the equation
motion (\ref{first equation A}), which is generated by the moving charge
itself. Namely, it represents the self-force acting on the moving charge.
It is widely known that, in the standard renormalization prescription,
such self-energy terms are absorbed into the redefinition of the
observed mass and charge of the charged particle, so that we shall simply
drop such self-energy terms in the following manipulation.) 

\vspace{2mm}
\noindent
The equal-time commutation relations for the relevant dynamical variables
are given as
\begin{eqnarray}
 &\,& \left[\, p_i, q_j \right] \ = \ - \,i \,\delta_{i j} , \\
 &\,& \left[ A_i (\bm{x}, t), E^\perp_j (\bm{x}^\prime, t) \right] \ = \ 
 - \,i \,\delta^{tr}_{i j} (\bm{x} - \bm{x}^\prime) ,
\end{eqnarray}
while other commutation relations are all zero. Here, 
$\delta^{tr}_{ij} (\bm{x} - \bm{x}^\prime)$ is the familiar transverse
delta function defined by
\begin{equation}
 \delta^{tr}_{i j} (\bm{x} - \bm{x}^\prime) \ = \ \int \,\frac{d^3 k}{(2 \,\pi)^3} \,\,
 e^{\,i \,\bm{k} \cdot ( \bm{x} - \bm{x}^\prime)} \,
 \left[ \delta_{i j} \ - \ \frac{k_i \,k_j}{\vert \bm{k} \vert^2} \right].
\end{equation}
We have confirmed that that the Heisenberg equations corresponding to 
the above Hamiltonian $H_A$ properly reproduces the equations of 
motion (\ref{allA}), which formally contains the self-interaction terms too.
Hereafter, the treatment based on $H_A$ will be called the scheme (A).

\vspace{2mm}
For our following discussion, it is critically important to divide the 
total electromagnetic field into two parts as
\begin{equation}
 A^\mu (x) \ = \ A^\mu_{ext} (x) \ + \ \tilde{A}^\mu (x).
\end{equation}
Here, $A^\mu_{ext} (x)$ is the part generated by the external 
current of the solenoid, while $\tilde{A}^\mu (x)$ is the part
generated by the moving charged particle. As already mentioned,
in view of the time-independence of the solenoid current, we can set
$A_{ext}^\mu (x) = (0, \bm{A}_{ext} (\bm{x}))$.
As fully discussed in Appendix \ref{AppA}, the general
form of $\bm{A}_{ext} (\bm{x})$ can be expressed in the form :
\begin{equation}
 \bm{A}_{ext} (\bm{x}) \ = \ 
 \bm{A}^{(S)}_{ext} (\bm{x}) \ + \ \nabla \chi (\bm{x})
 \ \equiv \ 
 \bm{A}^\perp_{ext} (\bm{x}) \ + \ 
 \bm{A}^\parallel_{ext} (\bm{x}), 
\end{equation}
with 
\begin{equation}
 \bm{A}^{(S)} (\bm{x}) \ = \ \frac{1}{4 \,\pi} \,\int \,d^3 x^\prime \,\,
 \frac{\bm{j}_{ext} (\bm{x})}{\vert \bm{x} - \bm{x}^\prime \vert} ,
\end{equation}
independently of the choice of gauge. Here, $\chi (\bm{x})$ can be
thought of as arbitrary non-singular scalar function.
In general, the vector potential $\tilde{\bm{A}} (\bm{x}, t)$ generated 
by the moving charge is also made up of the transverse and longitudinal parts.
In the following, to avoid unnecessary complexity, we can set 
$\tilde{\bm{A}}^\parallel (\bm{x}, t) = 0$, which corresponds to the complete 
Coulomb gauge fixing for the part $\tilde{\bm{A}} (\bm{x}, t)$.
Alternatively, we can think as if this longitudinal part
$\tilde{\bm{A}}^\parallel (\bm{x}, t)$ is absorbed into the longitudinal part 
$\bm{\bm{A}}_{ext}^\parallel (\bm{x})$ of the external vector potential.
In any case, the corresponding electromagnetic fields are represented as
\begin{eqnarray}
 \tilde{\bm{E}}_\perp (\bm{x}, t) &=& \bm{E}_\perp (\bm{x}, t) , \\
 \tilde{\bm{B}} (\bm{x}, t) &=& \nabla \times \tilde{\bm{A}} (\bm{x}, t), \\
 \bm{B}_{ext} (\bm{x}) &=& \nabla \times \bm{A}_{ext}  (\bm{x}). 
\end{eqnarray}
After the above separation, the equations of motion (\ref{allA}) 
can be transformed into the form
\begin{subequations} \label{allB}
 \begin{align}
 &\, m \,\frac{d^2 \bm{q}}{d t^2} \ = \ - \,\frac{\partial V}{\partial \bm{q}} \ + \ 
 e \,\left[ \tilde{\bm{E}}_\perp (\bm{q}, t) \ + \ \dot{\bm{q}} \times 
 (\tilde{\bm{B}} (\bm{q}, t) \ + \ \bm{B}_{ext} (\bm{q}) ) \right], 
 \label{first equation B} \\
 &\, \nabla \cdot \tilde{\bm{E}}_\perp (\bm{x}, t) \ = \ \tilde{j}^0 (\bm{x}, t) , 
 \label{second equation B} \\
 &\, \nabla \times \tilde{\bm{B}} (\bm{x}, t) \ = \ \tilde{\bm{j}} (\bm{x}, t) \ + \  
 \frac{\partial \tilde{\bm{E}}_\perp (\bm{x}, t)}{\partial t} \label{third equation B}.
 \end{align}
\end{subequations}
One can verify that the above equations of motion (\ref{allB})
can be derived from the following Lagrangian $L_B$ : 
\begin{eqnarray}
 L_B &=& \frac{1}{2} \,m \,\dot{\bm{q}}^2 \ - \ V (\bm{q}) \ - \ 
 \frac{1}{4} \,\int \,d^3 x \,\tilde{F}_{\mu \nu} \,\tilde{F}^{\mu \nu}
 \ - \ 
 \int \,d^3 x \,\,\tilde{j}_\mu (x) \,\left( \tilde{A}^\mu (x) \ + \ A^\mu_{ext} (x) \right),
\end{eqnarray}
with $\tilde{F}_{\mu \nu} = \partial_\mu \,\tilde{A}_\nu - \partial_\nu \,\tilde{A}_\mu$.
The Hamiltonian corresponding to this Lagrangian $L_B$ can be
constructed by  imposing the Coulomb gauge condition
$\nabla \cdot \tilde{\bm{A}} = 0$ for $\tilde{\bm{A}}$. It is given as
\begin{eqnarray}
 H_B &=& \frac{1}{2 \,m} \,\left( \bm{p} - e \,
 \left( \tilde{\bm{A}} (\bm{q}, t) + \bm{A}_{ext} (\bm{q}) \right) \right)^2 \ + \ V (\bm{q})
 \ + \ \int \,d^3 x \,\,
 \frac{1}{2} \,\left( \tilde{\bm{E}}^2_\perp + \tilde{\bm{B}}^2 \right) . \hspace{6mm}
 \label{HB}
\end{eqnarray}
The equal-time commutation relations for the relevant dynamical variables
are given by
\begin{eqnarray}
 &\,& \left[\, p_i, q_j \right] \ = \ - \,i \,\delta_{i j} , \\
 &\,& \left[ \tilde{A}^\perp_i (\bm{x}, t), \tilde{E}^\perp_j (\bm{x}^\prime, t) \right] \ = \ 
 - \,i \,\delta^{tr}_{i j} (\bm{x} - \bm{x}^\prime) ,
\end{eqnarray}
while other commutation relations are all zero. One can also verify that
the Heisenberg equations based on the Hamiltonian $H_B$ reproduce
the equations of motion (\ref{allB}).
Let us call this treatment based on $H_B$
the scheme (B). We point out that this scheme corresponds to the 
treatment described in the textbook of 
Cohen-Tannoudji et al. \cite{Cohen-Tannoudji1989}. 
(See Complement $\mbox{C}_{II}$ titled ``{\it Electrodynamics in the presence
of an external field}''.)
 
Since the two forms of Hamiltonians $H_A$ and $H_B$ are naturally expected to
describe the same dynamical system, which consists of the moving charge, the 
solenoid and the electromagnetic field, they must be related to each other. 
In fact, we can show that the two Hamiltonians
$H_A$ and $H_B$ basically coincide aside from physically irrelevant 
constant terms or the self-energy terms. 
To see it, we first notice that, with new field variables, 
$H_A$ can be expressed as
\begin{eqnarray}
 H_A \ &=& \ \frac{1}{2 \,m} \,\left[ \bm{p} - e \,\left( \tilde{\bm{A}} (\bm{q}, t) 
 + \bm{A}_{ext} (\bm{q}) \right) \right]^2 \ + \ V (\bm{q}) \nonumber \\
 &\,& \hspace{10mm} \ + \ 
 \int \,d^3 x \,\,\frac{1}{2} \,\left[ \tilde{\bm{E}}^2_\perp \ + \ 
 \left( \tilde{\bm{B}} + \bm{B}_{ext} \right)^2 \right] \ - \ 
 \int \,d^3 x \,\,\bm{j}_{ext} \cdot \left( \tilde{\bm{A}} + \bm{A}_{ext} \right) .
 \hspace{6mm}
 \label{HA_decomp}
\end{eqnarray}
Here, we note that
\begin{eqnarray}
 \frac{1}{2} \,\int \,d^3 x \,\left( \tilde{\bm{B}} \ + \ \bm{B}_{ext} \right)^2
 &=& \frac{1}{2} \,\int \,d^3 x \,\left( \tilde{\bm{B}}^2 \ + \ 
 2 \,\tilde{\bm{B}} \cdot \bm{B}_{ext}
 \ + \ \bm{B}^2_{ext} \right) \nonumber \\
 &=&  
 \frac{1}{2} \,\int \,d^3 x \,\,\tilde{\bm{B}}^2 \ + \ \int \,d^3 x \,\,
 \tilde{\bm{B}} \cdot \bm{B}_{ext} \ + \ \mbox{constant} ,
\end{eqnarray}
that is, the last term above represents the physically uninteresting self-energy
of the solenoid. On the other hand, the 2nd term of the above equation 
can be rewritten as 
\begin{eqnarray}
 \int \,d^3 x \,\tilde{\bm{B}} \cdot \bm{B}_{ext}
 &=& \int \,d^3 x \,\nabla \times \tilde{\bm{A}} \cdot \bm{B}_{ext} 
 \ = \ \int \,d^3 x \,\tilde{\bm{A}} \cdot \nabla \times \bm{B}_{ext} ,
\end{eqnarray}
with the use of integration by parts. The surface term of this integration
by parts can safely be neglected, since the magnetic field $\bm{B}_{ext}$ is
confined in a limited area, i.e. inside the solenoid. 
Then, using $\nabla \times \bm{B}_{ext} =
\bm{j}_{ext}$, we find that the following {\it nontrivial identity} holds :
\begin{equation}
 \int \,d^3 x \,\tilde{\bm{B}} \cdot \bm{B}_{ext} \ = \ 
 \int \,d^3 x \,\tilde{\bm{A}} \cdot \bm{j}_{ext} . \label{nt_relation}
\end{equation}
A little surprisingly, this piece of the magnetic interaction energy precisely cancels
the part $- \,\int \,d^3 x \,\bm{j}_{ext} \cdot \tilde{\bm{A}}$ in the last term
of (\ref{HA_decomp}). 
Besides, the part $- \,\int \,d^3 x \,\bm{j}_{ext} \cdot \bm{A}_{ext}$
in the last term of (\ref{HA_decomp}) reduces to a physically uninteresting 
self-interaction term.
As a consequence, we have succeeded to show that
\begin{equation}
 H_A \ = \ H_B \ + \ \mbox{constants} ,
\end{equation}
which confirms that the two Hamiltonian $H_A$ in (\ref{HA}) and $H_B$ 
in (\ref{HB}) coincide except for physically unimportant constant terms.

We are now ready to carry out a quantum mechanical analysis of 
the combined system of a solenoid and a moving charge
based on the two forms of Hamiltonian $H_A$ and $H_B$. 
The general strategy is that, in view of the smallness of the charge $e$, 
it is possible to neglect the second and higher order terms in $e$.

\subsection{Analysis based on the Hamiltonian $H_B$}\label{subsect3.1}

First, we analyze the system based on the Hamiltonian $H_B$.
Let us first split $H_B$ into the following pieces : 
\begin{equation}
 H_B \ = \ \frac{1}{2 \,m} \,\bm{p}^2 \ + \ V (\bm{q}) \ + \ H^\prime_B ,
\end{equation}
where
\begin{equation}
 H^\prime_B \ = \ \int d^3 x \,\,\frac{1}{2} \,
 \left( \tilde{\bm{E}}^2_\perp \ + \ \tilde{\bm{B}}^2 \right) \ - \ 
 \frac{e}{m} \,\,\bm{p} \cdot \left( \tilde{\bm{A}} (\bm{q}, t) \ + \ 
 \bm{A}_{ext} (\bm{q}) \right) \ + \ O (e^2) .  
\end{equation}
What should be quantized is the transverse part of $\tilde{\bm{A}} (\bm{x}, t)$,
and its longitudinal part can simply set to be zero or included in
the longitudinal part of $\bm{A}_{ext} (\bm{x})$, since the latter two
are just arbitrary c-number fields, anyway. 
Here, we use the standard free-field expansion
for the quantized field $\tilde{\bm{A}} (\bm{x}, t)$ in the Colomb gauge. 
(We recall that similar free field expansion was used also in
Saldanha's work but in the Lorenz gauge.)
This is justified, because our purpose here is to evaluate the interaction
energy between the solenoid and the moving charge based on the 1st
order perturbation theory in $e$. 
This gives
\begin{eqnarray}
 \tilde{\bm{A}} (\bm{x}, t) &=& \int \,\frac{d^3 k}{\sqrt{(2 \,\pi)^3}} \,\,
 \frac{1}{\sqrt{2 \,\omega}} \,\,\sum_{\lambda = 1}^2 \,\,
 \epsilon_{\bm{k}, \lambda} \,\left( \tilde{a}_{\bm{k}, \lambda} \,
 e^{\,i \,( \bm{k} \cdot \bm{x} - \omega \,t)}
 \ + \ \tilde{a}^\dagger_{\bm{k}, \lambda} \,
 e^{\,- \,i \,( \bm{k} \cdot \bm{x} - \omega \,t)} \right), \hspace{5mm}
 \label{Atilde_expansion}
\end{eqnarray}
with $\omega = \vert \bm{k} \vert$. Here, $\bm{\epsilon}_{\bm{k}, \lambda}$
and $\tilde{a} (\bm{k}, \lambda)$ respectively stand for the polarization
vector and annihilation operator of the photon with momentum $\bm{k}$
and polarization $\lambda = 1, 2$.
We denote the vacuum of the quanta $\tilde{a} (\bm{k}, \lambda)$ as
$\vert \tilde{0} \rangle$, i.e.
\begin{equation}
 \tilde{a} (\bm{k}, \lambda) \,\vert \tilde{0} \rangle \ = \ 0.
\end{equation}
Now, it is a textbook exercise to show that
\begin{equation}
 \int \,d^3 x \,\,\frac{1}{2} \,\,
 \left( \tilde{\bm{E}}^2_\perp \,+ \,\tilde{\bm{B}}^2 \right)
 \ = \ \sum_{\lambda = 1}^2 \,\int \,d^3 k \,\,\omega \,\,
 \tilde{a}^\dagger (\bm{k}, \lambda) \,\tilde{a} (\bm{k}, \lambda) .
\end{equation}
With the use of this relation together with the fact that
$\vert \tilde{0} \rangle$ is the vacuum of
the quantized electromagnetic field $\tilde{\bm{A}}$, we immediately
get
\begin{equation}
 \langle \tilde{0} \,\vert \,H^\prime_B \,\vert \,\tilde{0} \rangle \ = \ 
 - \,\frac{e}{m} \,\,\bm{p} \cdot \bm{A}_{ext} (\bm{q}) \ + \ O (e^2) .
\end{equation}
This gives
\begin{eqnarray}
 \langle \tilde{0} \,\vert \,H_B \,\vert \tilde{0} \rangle 
 &=& \frac{1}{2 \,m} \,\bm{p}^2
 \ + \ V (\bm{q}) \ - \ \frac{e}{m} \,\,\bm{p} \cdot \bm{A}_{ext} (\bm{q})
 \ + \ O (e^2). \hspace{6mm}
\end{eqnarray}
If we dare to revive a term of $O (e^2)$ so as to make clear the
gauge-covariance of the effective Hamiltonian, we may be able to
write as
\begin{eqnarray}
 \langle \tilde{0} \,\vert \,H_B \,\vert \tilde{0} \rangle 
 &=& \frac{1}{2 \,m} \,
 \left( \bm{p} \, - \, e \, \bm{A}_{ext} (\bm{q}) \right)^2 \ + \ V (\bm{q})
 \ + \ O (e^2) \ \equiv \ H_{particle}.\hspace{6mm}  \label{effective_HB}
\end{eqnarray}
It is clear that this form of effective Hamiltonian $H_{particle}$
provides us with the basis of the standard treatment of the 
AB-effect. In fact, under the gauge transformation of the external
gauge potential
\begin{equation}
 \bm{A}_{ext} (\bm{q}) \ \rightarrow \ \bm{A}^\prime_{ext} (\bm{q}) \ = \ 
 \bm{A}_{ext} (\bm{q}) \ + \ \nabla \chi (\bm{q}),
\end{equation}
$H_{particle}$ transform gauge-covariantly together with the wave function
of the charged particle, which is the essential mechanism of the AB-effect.
(For more rigorous treatment, see the later discussion based on the 
path integral formalism.)
This immediately raises the following question. Where in our Hamiltonian
$H_A$ or $H_B$ are the interaction energies as discussed by Boyer and/or
Saldanha and others hidden ?
An answer to this question is given in the next subsection.

\subsection{Analysis based on the Hamiltonian $H_A$}\label{subsect3.2} 

First, we split the Hamiltonian $H_A$ into the following pieces :
\begin{equation}
 H_A \ = \ \frac{1}{2 \,m} \,\bm{p}^2 \ + \ V (\bm{q}) \ + \ H^\prime_A ,
\end{equation}
with
\begin{eqnarray}
 H^\prime_A &=& \int \,d^3 x \,\,\frac{1}{2} \,\left( \bm{E}_\perp^2 + \bm{B}^2 \right)
 \ - \ \int \,d^3 x \,\,\bm{j}_{ext} \cdot \bm{A} \ - \ 
 \frac{e}{m} \,\bm{p} \cdot \bm{A} (\bm{q}) \ + \ O (e^2) \hspace{4mm}
 \nonumber \\
 &=& \hspace{3mm} H_{EM} \ + \ H_j \ + \ H_e \ + \ O (e^2),
\end{eqnarray}
where we have defined as
\begin{eqnarray}
 H_{EM} &\equiv& \int \,d^3 x \,\,\frac{1}{2} \,
 \left( \bm{E}^2_\perp \ + \ \bm{B}^2 \right) ,
 \\
 H_j &\equiv& - \,\int \,d^3 x \,\,\bm{j}_{ext} \cdot \bm{A} , \\
 H_e &\equiv& - \,\frac{e}{m} \,\,\bm{p} \cdot \bm{A} (\bm{q}) .
\end{eqnarray}
%

First, we carefully reexamine Boyer's analysis based on our 
Hamiltonian $H_A$. What he calculated is the
interaction energy between the magnetic field generated by the moving charge
and that generated by the solenoid current. The question is where we
can find such an interaction term in our Hamiltonian $H_A = 1 / (2 \,m) \,\bm{p}^2
+ V (\bm{q}) + H^\prime_A$ with
$H^\prime_A = H_{EM} + H_j + H_e + O (e^2)$ ?
To answer this question, we find it useful to treat
the Hamiltonian $H^\prime_A$ in the following manner.
That is, suppose that we look for the ground state of 
$H_{EM} + H_j + H_e$ using
the 1st order perturbation theory with respect to the small charge $e$.
It is given by
\begin{equation}
 \vert \tilde{0}_1 \rangle \ = \ \vert \tilde{0} \rangle \ + \ 
 \frac{1}{\tilde{E}_0 \, - \, ( H_{EM} \,+ \,H_j )} \,H_e \,\vert \tilde{0} \rangle ,
 \label{perturbation_e}
\end{equation}
where $\tilde{E}_0$ stands for the energy of the vacuum state $\vert \tilde{0} \rangle$.
With the use of $\vert \tilde{0}_1 \rangle$, let us calculate
the expectation value $\langle \tilde{0}_1 \,\vert \,H_{EM} + H_j + H_e \,
\vert \tilde{0}_1 \rangle$ up to the 1st order in $e$.
To this end, we first split $H_{EM}$ as follows : 
\begin{eqnarray}
 H_{EM} &=& \int \,d^3 x \,\,\frac{1}{2} \,\left\{ \tilde{\bm{E}}^2_\perp \ + \ 
 ( \tilde{\bm{B}} + \bm{B}_{ext} )^2 \right\} \nonumber \\
 &=& 
 \int \,d^3 x \,\,\frac{1}{2} \,\left( \tilde{\bm{E}}^2_\perp \ + \ \tilde{\bm{B}}^2 \right)
 \ + \ \int \,d^3 x \,\tilde{\bm{B}} \cdot \bm{\bm{B}}_{ext} \ + \ 
 \mbox{constant}  \nonumber \\
 &\equiv& \tilde{H}_{EM} \ + \
 \int \,d^3 x \,\tilde{\bm{B}} \cdot \bm{\bm{B}}_{ext} \ + \ 
 \mbox{constant}  .
\end{eqnarray}
Here, we have defined $\tilde{H}_{EM}$ by
\begin{equation}
 \tilde{H}_{EM} \ \equiv \ \int \,d^3 x \,\,\frac{1}{2} \,
 \left( \tilde{\bm{E}}^2_\perp \ + \ \tilde{\bm{B}}^2 \right).
\end{equation}
Obviously, this $\tilde{H}_{EM}$ can be expressed as
\begin{equation}
 \tilde{H}_{EM} \ = \ \sum_{\lambda = 1}^2 \,\int \,d^3 k \,\,\omega \,\,
 \tilde{a}^\dagger (\bm{k}, \lambda) \,\tilde{a} (\bm{k}, \lambda),
\end{equation}
so that it holds that $\langle \tilde{0} \,\vert \,\tilde{H}_{EM} \,
\vert \tilde{0} \rangle = 0$. Then, using this relation together with the fact 
that the difference between $\langle \tilde{0}_1 \,\vert \,\tilde{H}_{EM} \,
\vert \tilde{0}_1 \rangle$ and $\langle \tilde{0} \,\vert \,\tilde{H}_{EM} \,
\vert \tilde{0} \rangle = 0$ is $O (e^2)$, we have
\begin{equation}
 \langle \tilde{0}_1 \,\vert \,H_{EM} \,\vert \tilde{0}_1 \rangle \ = \ 
 \int \,d^3 x \,\,\langle \tilde{0}_1 \,\vert \,\tilde{\bm{B}} \,
 \vert \,\tilde{0}_1 \rangle \cdot \bm{B}_{ext} \ + \ O (e^2) .
\end{equation}
The expectation value $\langle \tilde{0}_1 \,\vert \,\tilde{\bm{B}} (\bm{x}) \,
\vert \tilde{0}_1 \rangle$ above can be evaluated as follows : 
\begin{eqnarray}
 \langle \tilde{0}_1 \,\vert \,\tilde{\bm{B}} (\bm{x}) \,
 \vert \tilde{0}_1 \rangle &=& 
 \langle \tilde{0}_1 \,\vert \,\nabla \times \tilde{\bm{A}} (\bm{x}) \,
 \vert \tilde{0}_1 \rangle \nonumber \\
 &=&  
 - \,\frac{e}{m} \,\langle \tilde{0} \,\vert \,\nabla \times 
 \tilde{\bm{A}} (\bm{x}) \,\,\frac{1}{\tilde{E}_0 \, - \, (H_{EM} \, + \,H_j)} \,\,
 \bm{p} \cdot \tilde{\bm{A}} (\bm{q}) \,\vert \tilde{0} \rangle \ + \ c.c.
 \hspace{6mm}
\end{eqnarray}
By using the form  for $H_{EM} + H_j$ given in Appendix.\ref{AppC} and 
the expansion (\ref{Atilde_expansion}) for $\tilde{\bm{A}} (\bm{x})$, 
it is not difficult to show that (see .\ref{AppD}, for derivation)
\begin{equation}
 \langle \tilde{0}_1 \,\vert \,\tilde{\bm{B}} (\bm{x}) \,
 \vert \tilde{0}_1 \rangle \ = \ - \,\frac{\bm{p}}{m} \times \nabla \,
 \frac{1}{4 \,\pi} \,\frac{e}{\vert \bm{x} - \bm{q} \vert} \ + \ O (e^2).
 \label{EV_Btilde}
\end{equation}
Interestingly, this gives the non-relativistic expression of the magnetic field 
generated by the charge moving with the velocity $\bm{v} = \bm{p} / m$, 
which was used in Boyer' analysis. 
The calculation hereafter proceeds just in the same manner as Boyer
did. We obtain
\begin{eqnarray}
 \langle \tilde{0}_1 \,\vert \,H_{EM} \,\vert \tilde{0}_1 \rangle 
 &=&- \,\frac{e}{m} \,\frac{1}{4 \,\pi} \,\int \,d^3 x \,
 \left( \bm{p} \times \nabla \,\frac{1}{\vert \bm{x} - \bm{q} \vert} \right)
 \cdot \bm{B}_{ext} (\bm{x}) \nonumber \\
 &=& - \,e \,\bm{v} \cdot \frac{1}{4 \,\pi} \,\int \,d^3 x \,
 \nabla \,\frac{1}{\vert \bm{x} - \bm{q} \vert} \times \bm{B}_{ext} (\bm{x})
 \nonumber \\
 &=& 
 e \,\bm{v} \cdot \frac{1}{4 \,\pi} \,\int \,d^3 x \,\,
 \frac{1}{\vert \bm{x} - \bm{q} \vert} \,\nabla \times \bm{B}_{ext} (\bm{x})
 \nonumber \\
 &=& e \,\bm{v} \cdot \frac{1}{4 \,\pi} \,\int \,d^3 x \,\,
 \frac{\bm{j}_{ext} (\bm{x})}{\vert \bm{x} - \bm{q} \vert} 
 \ = \ e \,\bm{v} \cdot \bm{A}^{(S)}_{ext} (\bm{q}) .
\end{eqnarray}
Here, $\bm{A}^{(S)}_{ext} (\bm{q})$ represents the axially-symmetric gauge
potential or the transverse part of $\bm{A}_{ext} (\bm{q})$.
Obviously, the expectation value of $H_{EM}$ derived above corresponds
to the interaction energy between the
magnetic field $\tilde{\bm{B}}$ generated by the moving charge and the
magnetic field $\bm{B}_{ext}$ generated by the external solenoid current
in Boyer's analysis. We recall, however, the fact 
that this interaction energy is different in sign from that of 
Saldanha. Now, we shall make clear the reason of this
perplexing observation. To see it, we first rewrite $H_j$ as
\begin{equation}
 H_j \ = \ - \,\int \,d^3 x \,\,\bm{j}_{ext} \cdot \bm{A} (\bm{x}) \ = \ 
 - \,\int \,d^3 x \,\,\bm{j}_{ext} (\bm{x}) \cdot \tilde{\bm{A}} (\bm{x})
 \ + \ \mbox{constant} .
\end{equation}
The next task is to calculate the expectation value
$\langle \tilde{0}_1 \,\vert \,H_j \,\vert \,\tilde{0}_1 \rangle$
by using (\ref{perturbation_e}).
Up to constants, we have 
\begin{eqnarray}
 \langle \tilde{0}_1 \,\vert \,H_j \,\vert \tilde{0}_1 \rangle \! = \!
 \langle \tilde{0} \,\vert \,H_j \,
 \frac{1}{\tilde{E}_0 - ( H_{EM} + H_j )} \,H_e + H_e \,
 \frac{1}{\tilde{E}_0 - ( H_{EM} + H_j )} \,H_j \,\vert \,\tilde{0} \rangle .
 \hspace{3mm}
\end{eqnarray}
Undoubtedly, this cross term of $H_j$ and $H_e$ corresponds to 
Saldanha's energy, i.e. the interaction energy between the solenoid
and the moving charge mediated by the exchange of a virtual photon.
It can be confirmed by the direct calculation as follows.
First, we get
\begin{equation}
 \langle \tilde{0}_1 \,\vert \,H_j \,\vert \tilde{0}_1 \rangle \ = \ 
 2 \,\,\frac{e}{m} \,\langle \tilde{0} \,\vert \,\int \,d^3 x \,\,
 \bm{j}_{ext} (\bm{x}) \cdot \tilde{\bm{A}} (\bm{x}) \,
 \frac{1}{\tilde{E}_0 - ( H_{EM} + H_J )} \,
 \bm{p} \cdot \tilde{\bm{A}} (\bm{q}) \,\vert \tilde{0} \rangle .
\end{equation}
Then, with the help of the manipulation
\begin{eqnarray}
 &\,& \langle \tilde{0} \,\vert \,\tilde{A}_i (\bm{x}) \,
 \frac{1}{\tilde{E}_0 \,-  \,( H_{EM} \, + \, H_j )} \,
 \bm{p} \cdot \tilde{\bm{A}} (\bm{q}) \,\vert \tilde{0} \rangle
 \nonumber \\
 &=& p_j \,\int \,d^3 k \,\frac{1}{(2 \,\pi)^3 \,2 \,\omega} \,
 \sum_{\lambda = 1,2} \,\frac{1}{- \,\omega} \,\,\,  
 \langle \tilde{0} \,\vert \,
 \epsilon_i (\bm{k}, \lambda) \,
 \left( \tilde{a} (\bm{k}, \lambda) \, e^{\,i \,\bm{k} \cdot \bm{x}} \ + \ 
 \tilde{a}^\dagger (\bm{k}, \lambda) \,e^{\,- \,i \,\bm{k} \cdot \bm{x}} \right) 
 \nonumber \\
 &\,& \hspace{49mm} \times \ 
 \epsilon_j (\bm{k}, \lambda) \,
 \left( \tilde{a} (\bm{k}, \lambda) \, e^{\,i \,\bm{k} \cdot \bm{x}} \ + \ 
 \tilde{a}^\dagger (\bm{k}, \lambda) \,e^{\,- \,i \,\bm{k} \cdot \bm{x}} \right)
 \,\vert \tilde{0} \rangle \hspace{8mm} \hspace{8mm} \nonumber \\
 &=& - \,p_j \,\int \,\frac{d^3 k}{(2 \,\pi)^3 \,2 \,\omega^2} \,
 \left( \delta_{ij} \ - \ \frac{k_i \,k_j}{\vert \bm{k} \vert^2} \right) \,
 e^{\,i \,\bm{k} \cdot (\bm{x} - \bm{q})} ,
 \end{eqnarray}
we obtain
\begin{eqnarray}
 \langle \tilde{0}_1 \,\vert \,H_j \,\vert \tilde{0}_1 \rangle
 &=& - \,\frac{e}{m} \,\int \,d^3 x \,\,j^i_{ext} (\bm{x}) \,p^j \,\,
 \int \,\frac{d^3 k}{(2 \,\pi)^3 \,\bm{k}^2} \,
 \left( \delta_{ij} \ - \ \frac{k_i \,k_j}{\vert \bm{k} \vert^2} \right) \,
 e^{\,i \,\bm{k} \cdot (\bm{x} - \bm{q})} .
\end{eqnarray}
Using integration by parts and the current conservation
$\nabla \cdot \bm{j}_{ext} (\bm{x}) = 0$, we can show that
the term $k_i \,k_j / \vert \bm{k} \vert^2$ drops out.
In this way, we eventually find that
\begin{eqnarray}
 \langle \tilde{0}_1 \,\vert \,H_j \,\vert \tilde{0}_1 \rangle &=&
 - \,\frac{e}{m} \,\int \,d^3 x \,\,\bm{p} \cdot \bm{j}_{ext} (\bm{x}) \,\,
 \frac{1}{4 \,\pi} \,\frac{1}{\vert \bm{x} - \bm{q} \vert} \nonumber \\
 &=& - \,\frac{e}{m} \,\,\bm{p} \cdot \frac{1}{4 \,\pi} \,
 \int \,d^3 x \,\frac{\bm{j}_{ext} (\bm{x})}{\vert \bm{x} - \bm{q} \vert} 
 \ = \ - \,\frac{e}{m} \,\,\bm{p} \cdot \bm{A}^{(S)}_{ext} (\bm{q}) .
\end{eqnarray}
This is precisely the interaction energy obtained by Saldanha.
Note that the sign of this energy is just opposite to the interaction
energy of Boyer.

\vspace{2mm}
Finally, the expectation value of $H_e$ becomes
\begin{eqnarray}
 \langle \tilde{0}_1 \,\vert \,H_e \,\vert \,\tilde{0}_1 \rangle &=& 
 - \,e \,\,\frac{\bm{p}}{m} \,\cdot \,
 \langle \tilde{0}_1 \,\vert \,\bm{A} (\bm{q}) \,\vert \,\tilde{0}_1 \rangle \nonumber \\
 &=& - \,e \,\,\frac{\bm{p}}{m} \,\cdot \,\langle \tilde{0} \,\vert \,\bm{A} (\bm{q}) \,
 \vert \,\tilde{0} \rangle \ + \ O(e^2) 
 \nonumber \\
 &=& - \,e \,\,\frac{\bm{p}}{m} \,\cdot \,\langle \tilde{0} \,\vert \,
 \tilde{\bm{A}} (\bm{q}) \,+ \,\bm{A}_{ext} (\bm{q}) \,
 \vert \,\tilde{0} \rangle \ + \ O(e^2) \nonumber \\
 &=& - \,e \,\,\frac{\bm{p}}{m} \cdot \bm{A}_{ext} (\bm{q})
 \ + \ O (e^2). \hspace{4mm}
\end{eqnarray}
It is very important to recognize that the vector potential appearing in this 
last equation is the full external potential 
$\bm{A}_{ext} (\bm{q}) = \bm{A}^{(S)}_{ext} (\bm{q})
\,+ \,\nabla \chi (\bm{q})$ {\it not} its gauge-invariant part
$\bm{A}^{(S)}_{ext} (\bm{q})$. 

\vspace{2mm}
To sum up, we find that
\begin{eqnarray}
 \langle \tilde{0}_1 \,\vert \,H_{EM} \,\vert \,\tilde{0}_1 \rangle 
 &=& + \,\frac{e}{m} \,\bm{p} \cdot \bm{A}^{(S)}_{ext} (\bm{q}) 
 \ = \ \Delta \epsilon \,(\mbox{Boyer}) , \\
 \langle \tilde{0}_1 \,\vert \,H_j \,\vert \,\tilde{0}_1 \rangle 
 &=& - \,\frac{e}{m} \,\bm{p} \cdot \bm{A}^{(S)}_{ext} (\bm{q})
 \ = \ \Delta \epsilon \,(\mbox{Saldanha}), \\
 \langle \tilde{0}_1 \,\vert \,H_e \,\vert \,\tilde{0}_1 \rangle 
 &=& - \,\frac{e}{m} \,\bm{p} \cdot \bm{A}_{ext} (\bm{q}) \ + \ 
 O (e^2).
\end{eqnarray}
This confirms that the interaction energy of Boyer and that of
Saldanha precisely cancel each other in our self-contained treatment 
of the system of the solenoid and the moving charge.   
As a consequence, we are eventually left with
\begin{equation}
 \langle \tilde{0}_1 \,\vert \,H_{EM} \ + \ H_j \ + \ H_e \,
 \vert \,\tilde{0}_1 \rangle \ = \  
 \langle \tilde{0}_1 \,\vert \,H_e \,\vert \tilde{0}_1 \rangle
 \ = \ 
 - \,e \,\,\frac{\bm{p}}{m} \cdot \bm{A}_{ext} (\bm{q}) 
 \ + \ O (e^2),
\end{equation}
within the approximation up to the first order in $e$.
In this way, we again find that
\begin{equation}
 \langle \tilde{0} \,\vert \,H_A \,\vert \,\tilde{0} \rangle \ = \ 
 \frac{1}{2 \,m} \,\left( \bm{p} \ - \ e \,\bm{A}_{ext} (\bm{q}) \right)^2 \ + \ 
 V (\bm{q}) \ + \ O (e^2) \ = \ H_{particle} .
\end{equation}
Remember that this effective Hamiltonian for a charged particle in the 
presence of the external magnetic field was more straightforwardly
obtained if we started with the Hamiltonian $H_B$. 
(See Eq.(\ref{effective_HB}).) However, somewhat redundant analysis 
as above was mandatory for explicitly showing the cancellation of the 
interaction energies of Boyer and of Saldanha and others.

\section{\label{sect4}Path integral formulation based on the Lagrangian $L_B$}

What can we learn after all from the analyses so far, concerning the 
question of the AB-phase shift for a non-closed path?
In a formal sense, the question can most clearly be answered by
using Feynman's path integral formalism based on the Lagrangian $L_B$
as will be explained below. First, by making use of the relation
\begin{eqnarray}
 \int \,d^3 x \,\,\tilde{j}_\mu (x) \,A^\mu_{ext} (x) &=&
 - \,\int \,d^3 x \,\,\tilde{\bm{j}} (\bm{x}, t) \cdot \bm{A}_{ext} (\bm{x}) \nonumber \\
 &=& - \,\int \,d^3 x \,\,e \,\dot{\bm{q}} \,\delta (\bm{x} - \bm{q} (t)) \cdot 
 \bm{A}_{ext} (\bm{x}) 
 \ = \ - \,e \,\dot{\bm{q}} \cdot \bm{A}_{ext} (\bm{x}) , 
 \ \ \ 
\end{eqnarray}
it is convenient to rewrite $L_B$ in the form
\begin{eqnarray}
 L_B &=& \frac{1}{2} \,m \,\dot{\bm{q}}^2 \ - \ V (\bm{q}) \ - \ 
 \frac{1}{4} \,\int \,d^3 x \,\,\tilde{F}_{\mu \nu} (x) \,\tilde{F}^{\mu \nu} (x)
 \ - \ \int \,d^3 x \,\,\tilde{j}_\mu (x) \,\tilde{A}^\mu (x) \ + \ 
 e \,\dot{\bm{q}} \cdot \bm{A}_{ext} (\bm{q}) .
\end{eqnarray}
Before explaining the path-integral formulation based on this form of 
Lagrangian $L_B$, we think it instructive to reconfirm the following 
features of  $L_B$. 
First, there is no piece in $L_B$ that represents the cross term of 
the magnetic field $\bm{B}^s$ generated by the solenoid and the magnetic
field $\tilde{\bm{B}}$ generated by the moving charge. 
This is obvious from the form of the electromagnetic energy
$- \frac{1}{4} \,\int \,d^3 x \,\tilde{F}_{\mu \nu} (x) \,\tilde{F}^{\mu \nu} (x)$,  
which contains only the field strength tensor $\tilde{F}_{\mu \nu} \equiv 
\partial_\mu \tilde{A}_\nu - \partial_\nu \tilde{A}_\mu$ with $\tilde{A}_\mu$
representing the piece of the electromagnetic potential generated by the 
moving charge.
Also noteworthy is the following feature.
First, the interaction term $- \,\int \,d^3 x \,\tilde{j}_\mu (x) \,\tilde{A}^\mu (x)$
between the solenoid current and the electromagnetic field contains
only the piece $\tilde{A}^\mu$ amongst the total electromagnetic field
$A^\mu = A_{ext}^\mu \,+ \,\tilde{A}^\mu$. On the other hand, the interaction
term $e \,\dot{\bm{q}} \cdot \bm{A}_{ext} (\bm{q})$ contains only the
piece $A_{ext}^\mu$ amongst the total electromagnetic field.
This means that, in $L_B$, there is no terms which mediate the interaction
between the solenoid and the moving charge by the exchange of a virtual
photon. 
From the above two observations, we can say again that the interaction
energy of Boyer and that of Saldanha are cancelling from the beginning
in the scheme based on the Lagrangian $L_B$. Although this
cancellation was already pointed out in the previous section, 
we nevertheless think it important to clearly reconfirm 
this fact again in order not to misunderstand the significance of the 
following path-integral analysis of the AB-phase shift based on the 
self-contained Lagrangian $L_B$. 

\vspace{2mm}
To move on, we split the Lagrangian $L_B$ into two pieces as
\begin{equation}
 L_B \ = \ L^0_B \ + \ L^{int}_B ,
\end{equation} 
with
\begin{eqnarray}
 L^0_B &=& \frac{1}{2} \,m \,\dot{\bm{q}}^2 \ - \ V (\bm{q}) \ - \ \frac{1}{4} \,
 \int \,d^3 x \,\,\tilde{F}_{\mu \nu} (x) \tilde{F}^{\mu \nu} (x)  \ - \ 
 \int \,d^3 x \,\,\tilde{j}_\mu (x) \,\tilde{A}^\mu (x) ,
\end{eqnarray}
and
\begin{equation} 
 L^{int}_B \ = \ e \,\dot{\bm{q}} \cdot \bm{A}_{ext} (\bm{q}) ,
\end{equation}

\vspace{2mm}
\noindent
The corresponding action is given by
\begin{equation}
 S_B \ \equiv \ \int_{t_i}^{t_f} \,L_B \,d t \ = \ S^0_B \ + \ S^{int}_B ,
\end{equation}
with
\begin{eqnarray}
 S^0_B &=& \int_{t_i}^{t_f} \,L^0_B \,d t , \label{SB_0} \\ 
 S^{int}_B &=& \int_{t_i}^{t_f} \,e \,\dot{\bm{q}} \cdot \bm{A}_{ext} (\bm{q}) \,d t .
 \label{SB_int}
\end{eqnarray}
We now show that the part $S^0_B$ is {\it gauge-invariant}, 
but $S^{int}_B$ is {\it not}.
First, note that, the term 
$\int \,d^4 x \,\tilde{F}_{\mu \nu} (x) \,\tilde{F}^{\mu \nu} (x)$
in $S^0_B$ is apparently gauge-invariant. 
We can also show that the term $\int \,d^4 x \,\tilde{j}_\mu (x) \,
\tilde{A}^\mu (x)$ in $S^0_B$ is gauge-invariant.
In fact, under the gauge transformation $\tilde{A}^\mu (x) \rightarrow
\tilde{A}^\mu (x) + \partial^\mu \tilde{\chi} (x)$, this term transforms as
\begin{eqnarray}
 \int \,d^4 x \,\tilde{j}_\mu (x) \,\tilde{A}^\mu (x) \ &\rightarrow& \  
 \int \,d^4 x \,\,\tilde{j}_\mu (x) \,\left( \tilde{A}^\mu (x) \ + \ 
 \partial^\mu \tilde{\chi} (x) \right) \nonumber \\
 &=& \int \,d^4 x \,\,\tilde{j}_\mu (x) \,\tilde{A}^\mu (x) \ - \ 
 \int \,d^4 x \,\,\partial^\mu \tilde{j}_\mu (x) \,\tilde{\chi} (x). \hspace{6mm}
\end{eqnarray}
By using the current conservation law $\partial^\mu \tilde{j}_\mu (x) = 0$,
the 2nd term vanishes, which shows that this term is certainly gauge-invariant.
Incidentally, the conservation of the current 
$\partial^\mu \tilde{j}_\mu (x) = 0$ can readily be verified from
the relations
\begin{eqnarray}
 \frac{\partial}{\partial t} \,\tilde{j}^0 (x) &=& 
 - \,e \,\dot{\bm{q}} \,\delta^\prime (\bm{x} - \bm{q} (t)), \\
 \nabla \cdot \tilde{\bm{j}} (x) &=& 
 + \, e \,\dot{\bm{q}} \,\delta^\prime (\bm{x} - \bm{q} (t)).
\end{eqnarray}

\vspace{2mm}
Now the path-integral formulation for the AB-phase based on the Lagrangian
$L_B$ goes just in the standard manner. 
We start with the expression for the
transition amplitude when the initial and final states are respectively specified 
by the wave functions of the charged particle
$\psi_i (\bm{q}) = \langle \bm{q} \vert \psi_i \rangle$ and
$\psi_f (\bm{q}) = \langle \bm{q} \vert \psi_f \rangle$. It is given by
\begin{eqnarray}
 \langle \psi_f, t_f \,\vert \,\psi_i, t_i \rangle \ = \ 
 \int \int d \bm{q}_f \,d \bm{q}_i \,\,\psi^*_f (\bm{q}_f (t_f)) \,
 K \left(\bm{q}_f, t_f \,; \,\bm{q}_i, t_i \right) \,
 \psi_i (\bm{q}_i (t_i)) . \label{Feynman_PI} \hspace{5mm}
\end{eqnarray}
To be more precise, the initial and final states are specified as
$\vert i \rangle = \vert \psi_i \rangle \,\vert \Phi^{(S)}_i \rangle \,\vert 0 \rangle$
and
$\vert f \rangle = \vert \psi_f \rangle \,\vert \Phi^{(S)}_f \rangle \,\vert 0 \rangle$,
where $\vert \psi_i \rangle$ represents the initial state of the charged particle,
$\vert \Phi^{(S)}_i \rangle$ the initial state of the solenoid, and $\vert 0 \rangle$
does the vacuum of the photon, and similarly for the final 
state \cite{Santos-Gonzalo1999}.
However, the solenoid always stays in the same stationary state, so that
the initial and final states are basically specified by those of the charge particle.
In (\ref{Feynman_PI}), $K (\bm{q}_f, t_f \,;\, \bm{q}_i, t_i) \equiv 
\langle \bm{q}_f, t_f \,\vert \,\bm{q}_i, t_i \rangle$ stands for the 
so-called Feynman kernel (or Feynman propagator) \cite{Feynman-Hibbs1965}, 
the path-integral representation of which is given as
\begin{eqnarray}
 &\,& \ \ K (\bm{q}_f, t_f \,;\, \bm{q}_i, t_i)  \nonumber \\
 &\,& \ = \int_{\bm{q}_i (t_i)}^{\bm{q}_f (t_f)} \,
 \,{\cal D} \bm{q} \,\int {\cal D} \tilde{A}^\mu \,\,
 \delta (\nabla \cdot \tilde{\bm{A}}) \,\exp \left[\, i S_B \right]
 \nonumber \\
 &\,& \ = \int_{\bm{q}_i (t_i)}^{\bm{q}_f (t_f)} \,\,{\cal D} \bm{q} \,
 \int {\cal D} \tilde{A}^\mu \,\,
 \delta (\nabla \cdot \tilde{\bm{A}}) \,
 \exp \,i \left[ \,S^0_B \, + \, e \,
 \int_{t_i}^{t_f} \,\dot{\bm{q}} \cdot \bm{A}_{ext} (\bm{q}) \,d t \,\right]
 \nonumber \\
 &\,& \ = \int_{\bm{q}_i (t_i)}^{\bm{q}_f (t_f)} {\cal D} \bm{q} \,
 \int {\cal D} \tilde{A}^\mu \,\,
 \delta (\nabla \cdot \tilde{\bm{A}}) \,
 \exp  \,i \left[ \,S^0_B \, + \, e \, 
 \int_{\bm{q} (t_i)}^{\bm{q} (t_f)} \,\bm{A}_{ext} (\bm{q}) \cdot d \bm{q} \,\right].
 \hspace{8mm} 
\end{eqnarray}
Here, we have inserted the gauge-fixing delta function corresponding to
the Coulomb gauge $\nabla \cdot \tilde{\bm{A}} = 0$ for the field
$\tilde{\bm{A}}$ in view of the
fact that the quantum theory of gauge field necessarily requires 
gauge-fixing. However, it is obvious that the Coulomb gauge is not mandatory
in the analysis here. Any gauge choice works equally well.

Important here is the gauge-transformation property of the Feynman kernel
$K (\bm{q}_f, t_f \,;\, \bm{q}_i, t_i)$. Although the part $S^0_B$ of the action is
invariant under the gauge transformation
$\bm{A}_{ext} (\bm{q}) \rightarrow \bm{A}_{ext} (\bm{q}) + \nabla \chi (\bm{q})$,
the part $S^{int}_B$ transforms as
 \begin{eqnarray}
  S^{int}_B &=& e \,\int_{\bm{q}_i}^{\bm{q}_f} \,\bm{A}_{ext} (\bm{q}) \cdot d \bm{q}
  \ \rightarrow \ e \,\int_{\bm{q}_i}^{\bm{q}_f} \,\,\left( \bm{A}_{ext} (\bm{q})
  \ + \ \nabla \chi (\bm{q}) \right) \cdot d \bm{q} \nonumber \\
  &=& 
  S^{int}_B \ + \ e \,\left[ \chi(\bm{q}_f) \ - \ \chi (\bm{q}_i) \right]. 
 \end{eqnarray}
Since the transition amplitude $\langle \psi_f, t_f \,\vert \,\psi_i, t_i \rangle$
should be gauge-invariant, it follows that the initial and final wave functions
of the charged particle must respectively transform as
\begin{eqnarray}
 \psi_i (\bm{q}_i) &\rightarrow& \psi^\prime_i (\bm{q}_i) \ = \ 
 e^{\,i \,e \,\chi (\bm{q}_i)} \,\psi_i (\bm{q}_i), \\
 \psi^*_f (\bm{q}_f) &\rightarrow& \psi^{\prime *}_f (\bm{q}_f) \ = \ 
 \psi^*_f (\bm{q}_f) \,e^{\,- \,i \,e \,\chi (\bm{q}_f)} .
\end{eqnarray}
This means that, when the charged particle travels from $\bm{q}_A$ to
$\bm{q}_B$ along a certain path $C$, it receives a phase change
given by
\begin{equation}
 \Delta \phi \ = \ e \,\int_{\bm{q}_A}^{\bm{q}_B} \,\bm{A}_{ext} (\bm{q})
 \cdot d \bm{q} \ + \ e \,\left[ \chi (\bm{q}_B) - \chi (\bm{q}_A) \right]
 \label{AB-phase_non-closed}
\end{equation}
which generally depends on the gauge function $\chi (\bm{q})$
representing the residual gauge degrees of freedom within the prescribed gauge. 
Naturally, in the special case where the initial and final positions of the charged
particle are the same, i.e. $\bm{q}_B = \bm{q}_A$, 
which corresponds to the case where the above path $C$ 
is closed, the gauge-dependent part $\chi (\bm{q}_B) - \chi (\bm{q}_A)$
vanishes and the phase change is just given by
\begin{equation}
 e \,\oint_C \,\bm{A}_{ext} (\bm{q}) \cdot d \bm{q}  \ = \ e \,\Phi ,
\end{equation}
where $\Phi$ is the total magnetic flux penetrating the closed path $C$. 
This is just the familiar AB-phase shift observed in the standard
interference experiment \cite{Tonomura1986,Osakabe1986}.

\section{\label{sect5}Summary and conclusion}

On the basis of the two forms of Hamiltonians $H_A$ and $H_B$, which are 
constructed so as to self-consistently describe the combined system of a 
solenoid and a moving charge, both of which are  interacting with the quantum 
electromagnetic field, we have carefully examined the recent claims by 
several authors that the
AB-phase shift of a charged particle along a non-closed path is
gauge-invariant and it can in principle be observed. 
The basic assumption taken for granted by these
authors is that the AB-phase shift is proportional to the change of
interaction energy between the solenoid and the charged particle along the
path of the moving charge.
In our treatment based on the Hamiltonian $H_A$, however, we find
that the change of energy obtained by Boyer and that obtained by
Saldanha and others are certainly gauge-invariant at least under
non-singular gauge transformations, but they precisely cancel 
each other out, which means that their claim loses its ground. 
This cancellation demonstrated based on the Hamiltonian $H_A$ 
becomes even more obvious if we start the
analysis from the physically equivalent Hamiltonian $H_B$. 
In fact, if one carefully looks
into the form of Hamiltonian $H_B$, one would take notice of the following fact.
First, there is no piece in $H_B$ that represents the cross term of the magnetic
field $\bm{B}^s$ generated by the solenoid and the magnetic field $\tilde{\bm{B}}$
generated by the moving charge. Remember that this cross term is
the origin of the interaction energy obtained by Boyer.
Second, in the Hamiltonian $H_B$, interactions terms that mediates the
interaction between the solenoid and the moving charge by the exchange
of a virtual photon are simply missing.
Remember that they are the origin of the interaction energy investigated
by Saldanha. Putting it another way, the interaction energy of Boyer and
that of Saldanha and others are cancelling from the outset in the scheme 
based on the Hamiltonian $H_B$. 
What is the origin of the AB-phase shift, then ?
To answer this question, we have carried out a path integral analysis based 
on the Lagrangian $L_B$ (corresponding to the Hamiltonian $H_B)$.
It was conformed that the AB-phase shift of a charged particle for a non-closed 
path is in fact a gauge-dependent quantity in contrast to the recent 
claims by several researchers. 

\vspace{3mm}
Concerning the debate on observability or nonobservability of the 
AB-phase shift for a non-closed path, worthy of special mention is 
a recent paper by Horvat et. al. \cite{Horvat2020}.
Just like Marletto and Vedral \cite{Marletto-Vedral2020},
to measure the AB-phase using only local operations and
classical communication, they proposed a measurement using
another reference electron called {\it ancillary} particle in addition
to a {\it primary} electron. However, they inherit the traditional
viewpoint that the AB-phase shift for a non-closed path is
a gauge-variant quantity.
In their argument, the time-variation of the magnetic
flux inside the solenoid during the measurement process plays an 
essential role. According to them, the  AB-phase is not acquired locally 
in the sense that only measurable 
quantities involved in this type of experiments correspond to 
the 4-vector potential around whole {\it space-time loops} and that the 
phase acquired along smaller portion of particle's path, i.e. the phase 
along a non-closed path, is not measurable. 
An advantage of this interpretation is that
their measurement does not contradict the gauge-invariance of 
the measured AB-phase. 
However, the necessity of the time-variation
of the magnetic flux appears to make their observable
deviate from the standard AB-phase obtained by the ordinary 
interference measurement. The reason is because,
if the magnetic flux inside the solenoid is varied in time,
an electric field is necessarily induced even outside the 
solenoid.
This is clearly different from the standard setting of the
Aharonov-Bohm effect, in which no electromagnetic field
exist outside the solenoid. Probably, measurement proposed
in \cite{Horvat2020} would rather be related to the so-called 
time-dependent Aharonov-Bohm effect \cite{Singleton-Vagenas2013}.

\begin{acknowledgments}
The author of the paper would  like to thank Akihisa Hayashi for
his valuable advises and enlightening discussions. Without his help, 
the paper would have never been accomplished.
The author would also like to thank the anonymous reviewer for many 
critical but constructive suggestions, which enhanced
the quality of the paper.
\end{acknowledgments}

\appendix

\section{\label{AppA}On the convergence of the integral given by Eq.(\ref{current_integral})}
As indicated in the main text, Boyer as well as Saldanha have implicitly assumed 
the convergence of Eq.(\ref{current_integral}) for a very long solenoid
without formally demonstrating it.
We point out that more rigorous argument was already given in the paper by Babiker and 
Lowdon \cite{Babiker-Lowdon1984}. 
They derived the following expression for the vector potential
of a finite solenoid of length $2 \,L$ and a radius of $R$, as measured in the $x$-$y$ plane.
The expression can be written as
\begin{equation}
   \bm{A}  \ = \  \frac{2 \,\Phi_{AB} \, \bm{e}_\phi}{\pi^2 \,R} \,\int_0^L \,
   \frac{(2 - k^2) \,\mathbb{K} (k) - 2 \,\mathbb{E} (k)}{k^2 \,\sqrt{(\rho + R)^2 + z^2}} \,d z,
   \label{A_finite_solenoid}
\end{equation}
where $\Phi_{AB}$ is the AB-flux corresponding to an infinitely long solenoid, 
$k^2 = 4 \,\pi \,R / [(\rho + R)^2 + z^2]$, and
$\mathbb{K} (k)$ and $\mathbb{E} (k)$ are the complete elliptical integrals of the first and 
second kind, respectively. Following them, 
it can be shown that along the regions $\rho \gg R$ and $\rho \ll R$, 
one can approximate Eq. (\ref{A_finite_solenoid}) as
\begin{equation}
 \bm{A} = \frac{\Phi_{AB} \,\bm{e}_\phi}{2 \,\pi \,\rho} \frac{L}{\sqrt{\rho^2 + L^2}} \ \
 (\rho \gg R), \hspace{6mm}
 \bm{A} = \frac{\Phi_{AB} \,\rho \,\bm{e}_\phi}{\pi \,R^2} 
 \frac{L}{\sqrt{R^2 + L^2} } \ \ (\rho \ll R). \label{A_finite_L}
\end{equation}
Clearly, these potentials do not correspond with those in the AB effect since we are dealing with a
finite-length solenoid, which is consistent with the fact that outside the solenoid the 
magnetic field is non-vanishing $\nabla \times \bm{A} \neq 0$. 
Accordingly, there would be a Lorentz force outside the solenoid acting on the
charge. However, from Eq. (\ref{A_finite_L}) it follows that such force would be negligible compared 
with other forces acting on the charge (e.g., a mechanical force) as $L$ increases. 
In this regard, Eq. (\ref{A_finite_L}) gives a qualitative
idea of what is meant by a ``very long solenoid". 
In fact, in the limit of an infinitely long solenoid ($L \rightarrow \infty)$ we obtain the 
expected result
\begin{equation}
 \lim_{L \rightarrow \infty} \,\bm{A} = \frac{\Phi_{AB} \,\bm{e}_\phi}{2 \,\pi \,\rho} \ \ 
 (\rho > R), \hspace{6mm}
  \lim_{L \rightarrow \infty} \,\bm{A} = \frac{\Phi_{AB} \,\rho \,\bm{e}_\phi}{\pi \,R^2} \ \ 
 (\rho < R).
\end{equation}
On the other hand, when $L \gg \rho$ and $L \gg R$, we obtain the approximate expressions
\begin{equation}
 \bm{A} \vert_{L \gg \rho} \simeq \frac{\Phi_{AB} \,\bm{e}_\phi}{2 \,\pi \rho} \ \
  (\rho \gg R), \hspace{6mm}
  \bm{A} \vert_{L \gg R} \simeq \frac{\Phi_{AB} \,\rho \,\bm{e}_\phi}{\pi \,R^2} \ \
  (\rho \ll R). 
\end{equation}
For example, if $L = 10^3 \,\rho$ and $\rho = 10^3 \,R$, then outside the solenoid the vector 
potential takes the form
\begin{equation}
 \bm{A} = 0.99 \times \left( \frac{\Phi_{AB} \,\bm{e}_\phi}{2 \,\pi \,\rho} \right) \simeq
 \frac{\Phi_{AB} \,\bm{e}_\phi}{2 \,\pi \rho} ,
\end{equation}
This clarifies in what sense the expression of the vector potential given
in Boyer's paper is justified.

\vspace{2mm}
\section{\label{AppB}On gauge-choice independence of Boyer's interaction energy}
Here, we address to the question whether the interaction energy of
Boyer has truly a gauge-invariant meaning {\it beyond} the Coulomb gauge.
To this end, we ask ourselves the following question : 

\begin{itemize}
\item What is the most general form of vector potential generated
by the steady current of an extremely long solenoid ?
\end{itemize}

\noindent
To answer this question, we start with one of the (time-dependent) Maxwell 
equation, 
\begin{equation}
 \nabla \times \bm{B} (\bm{x}, t) \ - \ 
 \frac{\partial \bm{E} (\bm{x}, t)}{\partial t} \ = \ \bm{j} (\bm{x}, t).
\end{equation}
If one introduces the electromagnetic
potentials $\phi$ and $\bm{A}$ by the familiar relations 
$\bm{B} = \nabla \times \bm{A}$ and
$\bm{E} = - \,\nabla \phi - \frac{\partial \bm{A}}{\partial t}$,
the above Maxwell equation becomes
\begin{equation}
 \nabla \left( \nabla \cdot \bm{A} (\bm{x}, t) \ + \,
 \frac{\partial \phi (\bm{x}, t)}{\partial t} \right)
 \ - \ \Delta \bm{A} (\bm{x}, t) \ - \ 
 \frac{\partial^2 \bm{A} (\bm{x}, t)}{\partial t^2} \ = \ 
 \bm{j} (\bm{x}, t) ,
\end{equation}
Since the magnetic field
inside the solenoid is thought to be generated by a {\it steady}
(time-independent) surface current $\bm{j} (\bm{x})$ of the solenoid, 
we can think that $\phi$ and $\bm{A}$ are both time-independent
without loss of generality. 
Under such circumstances, the above equation reduces to 
\begin{equation}
 \Delta \bm{A} (\bm{x}) \ - \,\nabla \,( \nabla \cdot \bm{A} (\bm{x})) 
 \ = \ - \,\bm{j} (\bm{x}). \label{General_A}
\end{equation}
independently of the gauge condition for $\bm{A} (\bm{x})$.
The solution of the above equation can most easily be found if we
impose the Coulomb gauge condition $\nabla \cdot \bm{A}^{(C)} (\bm{x}) = 0$
for the vector potential. In fact, under this condition, the above
equation reduces to the familiar Poisson equation \cite{Shadowitz1975} :
\begin{equation}
 \Delta \bm{A}^{(C)} (\bm{x}) \ = \ - \,\bm{j} (\bm{x}) , \label{A_Poisson_eq}
\end{equation}

\vspace{2mm}
\noindent
Here, $\bm{A}^{(C)} (\bm{x}) $ stands for the vector potential satisfying 
the above Poisson equation together with the Coulomb gauge condition.
As described in standard textbooks, the general
solution to this Poisson equation is given as
\begin{equation}
 \bm{A}^{(C)} (\bm{x}) \ = \ \bm{A}^{(S)} (\bm{x}) \ + \ \nabla \chi^{(H)} (\bm{x}),
 \label{Cgauge_A}
\end{equation}
where
$\bm{A}^{(S)} (\bm{x})$ is given by \footnote{The superscript $(S)$ of 
$\bm{A}^{(S)} (\bm{x})$ has
double meanings. On the one hand, it means the Special solution of
the above Poisson equation (\ref{A_Poisson_eq}). 
On the other hand, it means that
the vector potential $\bm{A}^{(S)} (\bm{x})$ is uniquely given as a definite 
convolution integral of the source current $\bm{j} (\bm{x})$.}
\begin{equation}
 \bm{A}^{(S)} (\bm{x}) \ = \ \frac{1}{4 \,\pi} \,\int \,d^3 x^\prime \,\,
 \frac{\bm{j} (\bm{x}^\prime)}{\vert \bm{x} - \bm{x}^\prime \vert}.
\end{equation}
while $\chi^{(H)} (\bm{x})$ is any scalar function satisfying the
Laplace equation $\Delta \chi^{(H)} (\bm{x}) = 0$.
The last condition $\Delta \chi^{(H)} = 0$ follows, because $\bm{A}^{(C)}$
must satisfy the Coulomb gauge condition, which dictates that 
\begin{equation}
 0 \ = \ \nabla \cdot \bm{A}^{(C)} \ = \ 
 \nabla \cdot \bm{A}^{(S)} \ + \ \Delta \chi^{(H)}
 \ = \ \Delta \chi^{(H)}.
\end{equation}
Here, use has been made of the relation $\nabla \cdot \bm{A}^{(S)} = 0$,
which can be easily verified by using the conservation law
$\nabla \cdot \bm{j} (\bm{x}) = 0$ for the steady current.
Unfortunately, the above form of the vector potential holds only in
the Coulomb gauge. In fact, $\bm{A}^{(C)}$ does not satisfy more general
field equation (\ref{General_A}) for the vector potential.
A question is whether one can find more general solution that satisfies 
Eq.(\ref{General_A}), which holds in arbitrary gauge.
Actually, this is not so difficult. In fact, one can easily confirm that
the following form of the vector potential satisfies the 
equation (\ref{General_A}),
\begin{equation}
 \bm{A} (\bm{x}) \ = \ \bm{A}^{(S)} (\bm{x}) \ + \ \nabla \chi (\bm{x}),
 \label{General_form_A}
\end{equation}
where $\chi (\bm{x})$ is an arbitrary scalar function not restricted to
a Harmonic function. In fact, it holds that
\begin{eqnarray}
 \Delta \bm{A} (\bm{x}) \, - \, \nabla (\nabla \cdot \bm{A} (\bm{x}))
 \ &=& \ \Delta \,(\bm{A}^{(S)} (\bm{x}) + \nabla \chi (\bm{x})) \, - \, 
 \nabla \,(\nabla \cdot \bm{A}^{(S)} (\bm{x}) + \Delta \chi (\bm{x})) \nonumber \\
 \ &=& \ \Delta \,\bm{A}^{(S)} (\bm{x}) \ = \ \Delta \,\,\frac{1}{4 \,\pi} \,
 \int \,d^3 x^\prime \,\,\frac{\bm{j} (\bm{x}^\prime)}
 {\vert \bm{x} - \bm{x}^\prime \vert} \nonumber \\
 \ &=& \ \frac{1}{4 \,\pi} \,\int \,d^3 x^\prime \,\,( - \,4 \,\pi) \,\,
 \delta (\bm{x} - \bm{x}^\prime) \,\bm{j} (\bm{x}^\prime)
 \ = \ - \,\bm{j} (\bm{x}).
\end{eqnarray}
We thus find that the vector potetial generated by the steady current of 
a very long solenoid can always be expressed in the form (\ref{General_form_A}) 
{\it independently} of the gauge choice. 
The part $\bm{A}^{(S)} (\bm{x})$ looks clearly gauge
invariant since it is expressed as a definite
convolution integral of the source current.
On the other hand, the part $\nabla \chi (\bm{x})$, which carries the gauge 
degrees of freedom of $\bm{A} (\bm{x})$, is totally arbitrary.
Here, a natural guess is that $\bm{A}^{(S)} (\bm{x})$ and $\nabla \chi (\bm{x})$
can respectively be identified with the transverse part $\bm{A}_\perp (\bm{x})$
and the longitudinal part $\bm{A}_\parallel (\bm{x})$ of the vector 
potential \footnote{We point out that,
even restricting to regular gauge transformations, there is some special 
physical problem in which the transverse component of the vector potential 
cannot be uniquely determined. This happens, for example, in the famous Landau
problem that handles the motion of an electron in a magnetic
field uniformly spreading over the whole 2-dimensional plane.
In such special circumstances, there exist plural forms of vector
potential satisfying the transverse condition. In fact, one can easily confirm
that the vector potentials with the three typical gauge choices, i.e. the symmetric 
gauge and the two Landau gauges, all satisfy the transverse 
condition \cite{WKZ2018},\cite{WH2022}.} .

\vspace{2mm}
\noindent
Unfortunately. there still remains a delicate problem in this identification.
In fact, since the function $\chi$ is an arbitrary scalar function, one can
in principle consider the following singular or multi-valued gauge function, 
\begin{equation}
  \chi (\bm{x}) \ = \ - \,\Phi \,\phi \ = \ - \,\Phi \,\arctan \left( \frac{y}{x} \right) ,
  \label{singular_chi}
\end{equation}
with $\bm{x} = (\bm{\rho}, z) = (\rho, \phi, z)$. 
As shown below, the rotation of $\nabla \chi$ does not vanish, i.e.
$\nabla \times (\nabla \chi) \neq 0$.  Rather, it can be shown that $\nabla \chi$ 
satisfies the transverse condition, $\nabla \cdot (\nabla \chi) = 0$.
This means that the transverse-longitudinal decomposition, or more
precisely, the identification of the transverse component of the vector
potential is not unique, once the multi-valued gauge function as
above is allowed.

\vspace{2mm}
\noindent
Still, it seems to us that such a multi-valued gauge transformation is
a little unnatural as well as artificial  from a physical viewpoint.
The reason will be explained below. We already know that,
assuming an appropriate limiting procedure, the vector potential defined by
\[
 \bm{A}^{(S)} (\bm{x}) \ = \ \int \,\frac{\bm{J}_{ext} (\bm{x}^\prime)}{\vert \bm{x} - \bm{x}^\prime \vert}
 \,d^3 x^\prime
\]
with the surface current of the solenoid given by
\[
 \bm{J}_{ext} (\bm{x}) \ = \ B \,\delta (\rho - R) \,\bm{e}_\phi
\]
takes the following axially-symmetric form
\[
 \bm{A}^{(S)} (\bm{x}) \ = \ \frac{\Phi}{2 \,\pi} \,
 \left[ \,\frac{\rho \,\theta (R - \rho)}{R^2} \ - \ \frac{\theta (\rho - R)}{\rho} \,\right]
 \,\bm{e}_\phi ,
 \]
where $\bm{x} = (\bm{\rho}, z) = (\rho, \phi, z)$. Suppose that we define a
new vector potential $\bm{A}^\prime (\bm{x})$ through the multi-valued
gauge transformation
\[
 \bm{A}^\prime (\bm{x}) \ = \ \bm{A}^{(S)} (\bm{x}) \ + \ \nabla \chi (\bm{x}),
\]
with the singular gauge function $\chi$ given by Eq.(\ref{singular_chi}).
%
%
Since 
$\partial^2 \chi / \partial x \,\partial y \neq \partial^2 \chi / \partial y \,\partial x$
for the singular gauge function $\chi$,
$\nabla \times (\nabla \chi)$ does not vanish. In fact, we find that
\[
 \nabla \times (\nabla \chi) \ = \ - \,\Phi \,\delta (x) \,\delta (y) \,\bm{e}_z
\] 
Introducing the notation
\[
 \bm{A}^\prime (\bm{x}) \ = \ \bm{A}^{(S)} (\bm{x}) \ + \ \nabla \chi (\bm{x})
 \ \equiv \ \bm{A}^{(S)} (\bm{x}) \ + \ \bm{A}_{string} (\bm{x}),
\]
we therefore find that
\begin{eqnarray*}
 \bm{B}^\prime (\bm{x}) &=& \nabla \times \bm{A}^\prime (\bm{x}) 
 \ = \ \nabla \times \bm{A}^{(S)} (\bm{x}) \ + \ \nabla \times \bm{A}_{string} (\bm{x}) \\
 &=& B \,\theta (R - \rho) \,\bm{e}_z  \ - \ 
 \Phi \,\delta (x) \,\delta (y) \,\bm{e}_z \nonumber \\
 &=& \bm{B}^{(S)} (\bm{x}) \ + \ \bm{B}_{string} (\bm{x}) .
 \end{eqnarray*}
Note that the total magnetic flux penetrating the solenoid vanishes, since
\begin{eqnarray*}
 \Phi^\prime &\equiv& \int \,\bm{B}^{(S)} (\bm{x}) \cdot d \bm{S} \ + \ 
 \int \,\bm{B}_{string} (\bm{x}) \cdot d \bm{S} 
 \ = \ \pi \,R^2 \ - \ \Phi \ = \ 0.
\end{eqnarray*}
Also widely-known is the fact that the vector potential outside the solenoid
can be completely eliminated by the above multi-valued gauge transformation,
i.e.
\begin{equation}
 \bm{A}^\prime (\bm{x}) \ = \ 0 \ \ \ \mbox{for} \ \ \ \rho > R
\end{equation}
Based on this last observation, Bocchieri and Loisinger once erroneously
claimed that the Aharonov-Bohm effect does not exist \cite{BL1978}.
As shown by several researchers, the AB-effect remains to exist
even after such a multi-valued gauge transformation, if one properly
takes account of the change of the $2 \,\pi$ periodic boundary condition
for the electron wave function \cite{Kreizschmar1965,BB1983,Miyazawa2}. 

\vspace{3mm}
\noindent
Although mathematically allowable, however, we point out that very peculiar
nature of the above-mentioned multi-valued gauge transformation.
To explain it, let us recall the basic definition of the 
familiar transverse-longitudinal
decomposition of the vector field $\bm{F}$ is given as
\[
 \bm{F} \ = \ \bm{F}_\perp \ + \ \bm{F}_\parallel.
\]
where the transverse component $\bm{F}_\perp$ and the longitudinal component
$\bm{F}_\parallel$ are respectively demanded to satisfy the following transverse
condition  and the longitudinal condition :
\[
 \nabla \cdot \bm{F}_\perp \ = \ 0, \ \ \ \nabla \times \bm{F}_\parallel \ = \ 0.
\]
Note that, since $\nabla \times \bm{A}_{string} \neq 0$, the part 
$\bm{A}_{string} = \nabla \chi$
with $\chi = - \,\Phi \,\arctan (y / x)$ is not categorized into a longitudinal 
component any more, but it is a transverse component. 
In fact, one can verify that it satisfies the transverse condition,
\[
 \nabla \cdot \bm{A}_{string} (\bm{x}) \ = \ 0.
\]

\vspace{1mm}
\noindent
This means that the transverse part of the vector potential is neither gauge invariant
nor unique, if multi-valued gauge transformation is allowed. However, we continue to
argue unphysical nature of such multi-valued gauge transformation.
Suppose that we calculate rotation of the magnetic field $\bm{B}^{(S)} (\bm{x})$
and $\bm{B}_{string} (\bm{x})$, the sum of which gives the new magnetic field
$\bm{B}^\prime (\bm{x}) = \bm{B}^{(S)} (\bm{x}) + \bm{B}_{string} (\bm{x})$
after singular gauge transformation. We find that
\begin{eqnarray*}
 \nabla \times \bm{B}^{(S)} (\bm{x}) &=& B \,\delta (\rho - R)  \,\bm{e}_\phi
 \ = \ \bm{J}_{ext} (\bm{x}), \\
 \nabla \times \bm{B}_{string} (\bm{x}) &=& \Phi \,\left( \,
 \delta (x) \,\delta^\prime (y) \,\bm{e}_x \ - \ 
 \delta^\prime (x) \,\delta (y) \,\bm{e}_y \,\right) \ \equiv \ \bm{J}_{string} (\bm{x}).
\end{eqnarray*}
This means that the new magnetic field $\bm{B}^\prime (\bm{x})$ satisfies the 
following equation
\[
 \nabla \times \bm{B}^\prime (\bm{x}) \ = \ \bm{J}_{ext} (\bm{x}) \ + \ 
 \bm{J}_{string} (\bm{x}) ,
\]
which is different from the original Maxwell equation
\[
 \nabla \times \bm{B}^{(S)} (\bm{x}) \ = \ \bm{J}_{ext} (\bm{x}).
\]

\vspace{2mm}
\noindent
In this way, we now realize that the multi-valued gauge transformation changes the 
magnetic field distribution inside the solenoid, and formally this change of the magnetic
field distribution is thought to be generated by a peculiar effective source current 
$\bm{J}_{string} (\bm{x})$ induced by
the above singular gauge transformation.  
A logical consequence of this consideration is that it might also affect
Boyer's interaction energy between the solenoid and the charged particle. 

\vspace{3mm}
\noindent
To confirm it, let us first remember the expression of Boyer's interaction energy
written as
\[
 \Delta \varepsilon (Boyer) \ = \ \frac{1}{4 \,\pi} \,\int \,d^3 x \,
 \bm{B}^s \cdot (\bm{v} \times \bm{E}^\prime) \ = \ 
 \bm{v} \cdot \frac{1}{4 \,\pi} \,\int \,d^3 x \,\bm{E}^\prime \times \bm{B}^s .
\]
with
\begin{eqnarray*}
 \bm{E}^\prime (\bm{x}, t) &=& - \,\nabla \,\frac{e}{\vert \bm{x} - \bm{x}^\prime \vert}, 
 \hspace{6mm} 
 \bm{B}^s (\bm{x}) \ = \ B \,\theta (R - \rho) \,\bm{e}_z .
\end{eqnarray*}
As already pointed out, under the multi-valued gauge transformation, 
the magnetic field distribution changes as
\[
 \bm{B}^s (\bm{x}) \ \rightarrow \ \bm{B}^s (\bm{x}) \ + \ \bm{B}_{string} (\bm{x}).
\]

\vspace{1mm}
\noindent
This means that Boyer's energy also changes as
\[
 \Delta \varepsilon (Boyer) \ \rightarrow \ \Delta \varepsilon^\prime (Boyer) \ = \ 
 \Delta \varepsilon (Boyer) \ + \ \Delta \varepsilon (string),
\]
where
\[
 \Delta \varepsilon (string) \ = \ 
 \bm{v} \cdot \frac{1}{4 \,\pi} \,\int \,d^3 x \,\bm{E}^\prime \times \bm{B}_{string} .
\]
After some algebra, we can show that
\[
 \Delta \varepsilon (string) \ = \ - \,e \,\bm{v} \cdot \left( \,\frac{\Phi}{2 \,\pi} \,
 \frac{1}{\rho^\prime} \right) \,\bm{e}_{\phi^\prime} .
\]
Comparing this with the expression of $\Delta \varepsilon (Boyer)$ given as
\[
 \Delta \varepsilon (Boyer) \ = \ e \,\bm{v} \cdot \frac{\Phi}{2 \,\pi} \,
 \left[ \frac{\rho^\prime \,\theta (R - \rho^\prime)}{R^2} \ + \ 
 \frac{\theta (\rho^\prime - R)}{\rho^\prime} \right] \,\bm{e}_{\phi^\prime} ,
\]
we find that
\[
 \Delta \varepsilon^\prime (Boyer) \ = \ \Delta \varepsilon (Boyer) \ + \ 
 \Delta \varepsilon (string) \ = \ 0. 
\]
outside the solenoid, i.e. for $\rho^\prime > R$, which is the region 
where the charged particle is making a rectilinear motion.
That is, Saldanha's interaction energy becomes zero after the multi-valued
gauge transformation.
As is discussed in the main text, the expression of Saldanha's energy
is just the same as Boyer's energy except for the sign difference, so that
it is clear that the Saldanha's energy also becomes zero after the
multi-valued gauge transformation. This means that the cancellation between the
energies of Boyer and of Saldanha  remains intact even after the multi-valued
gauge transformation.

\vspace{2mm}
\noindent
In this way,  we realize that 
that the interaction energies of Boyer (and also of Saldanha) is not gauge-invariant,
if a multi-valued gauge transformation is allowed.
However, remember that the multi-valued gauge transformation induces
peculiar effective current and it changes the form of the Maxwell equation for the
magnetic field. Then, even though such a multi-valued gauge transformation
is mathematically allowed, we feel it very unnatural from a physical viewpoint.
On the other hand, if we confine to regular gauge transformation, we can say that the
transverse part of the gauge potential is gauge-invariant as well as unique, and the 
energies of Boyer and Saldanha are certainly expressed with this transverse component,
even though these two energies are destined to cancel each other out.

\section{\label{AppC}On the identity for $H_{EM} + H_j$}

With the use of Eqs. (90) and (97), we get
\begin{eqnarray}
 H_{EM} + H_j &=& \tilde{H}_{EM} + \int d^3 x \,\,
 \tilde{\bm{B}} (\bm{x}) \cdot \bm{B}_{ext} (\bm{x}) 
 \ - \ 
 \int d^3 x \,\,\tilde{\bm{A}} (\bm{x}) \cdot \bm{j}_{ext} (\bm{x}) + 
 \mbox{constant} .
\end{eqnarray}
Note that the 2nd and 3rd terms on the r.h.s. of the above equation cancel
out owing to the nontrivial identity (74) in the text. Thus, we have
\begin{eqnarray}
 H_{EM} + H_j &=& \tilde{H}_{EM} + \mbox{constant}
 \ = \ \sum_{\lambda = 1}^2 \,\int d^3 k \,\,\omega \,
 \tilde{a}^\dagger (\bm{k}, \lambda) \,\tilde{a} (\bm{k}, \lambda)
 + \mbox{constant} .
\end{eqnarray}
This means that the ground state of $H_{EM} + H_j$ is the vacuum
$\vert \tilde{0} \rangle$ of the quanta $\tilde{a} (\bm{k}, \lambda)$,
so that it holds that
\begin{equation}
 \left( H_{EM} + H_j \right) \,\vert \tilde{0} \rangle \ = \ 
 \tilde{E}_0 \,\vert \tilde{0} \rangle .
\end{equation}
Here, $\tilde{E}_0$ represents the ground-state energy of $H_{EM} + H_j$,
which may in principle contain self-energies of the solenoid and
the moving charge.

\vspace{3mm}
\section{\label{AppC}Proof of Eq.(\ref{EV_Btilde}).}

Up to $O (e^2)$, we have 
\begin{eqnarray}
 &\,& \langle \tilde{0}_1 \vert \,\tilde{\bm{B}} (\bm{x}) \vert \,
 \tilde{0}_1 \rangle
 \ = \ - \ \frac{e}{m} \,\langle \tilde{0} \,\vert \,
 \nabla \times \tilde{\bm{A}} (\bm{x}) \,
 \frac{1}{\tilde{E}_0 - (H_{EM} + H_j )} \,
 \bm{p} \cdot \tilde{\bm{A}} (\bm{q}) \,\vert \tilde{0} \rangle \ + \ c.c
\end{eqnarray}
In view of the Fourier expansion of $\tilde{\bm{A}} (\bm{x})$, this gives
\begin{eqnarray}
 &\,& \langle \tilde{0}_1 \vert \,\tilde{\bm{B}} (\bm{x}) \vert \,
 \tilde{0}_1 \rangle 
 \ = \ \ - \,2 \,\,\frac{e}{m} \,\int \frac{d^3 k}{(2 \,\pi)^3} \,\sum_{\lambda = 1}^2 \,\,
 \frac{\langle \tilde{0} \,\vert \,\nabla \times \tilde{\bm{A}} (\bm{x}) \,
 \vert \bm{k}, \lambda \rangle \,\langle \bm{k}, \lambda \,\vert \,
 \bm{p} \cdot \tilde{\bm{A}} (\bm{q}) \,\vert \tilde{0} \rangle}
 {- \,\omega} \,\, \frac{1}{2 \,\omega} . \hspace{5mm}
\end{eqnarray}
Now, we proceed as
\begin{eqnarray}
 &\,& \int \frac{d^3 k}{(2 \,\pi)^3} \,\frac{1}{- \,\omega^2} \,
 \sum_{\lambda = 1}^2 \,\langle \tilde{0} \,\vert \,
 \nabla \times \tilde{\bm{A}} (\bm{x}) \,\vert \bm{k}, \lambda \rangle \,
 \langle \bm{k}, \lambda \,\vert \bm{p} \cdot \tilde{\bm{A}} (\bm{q}) \,
 \tilde{0} \rangle \nonumber \\
 &\,& = - \, 2 \,i \,\int \frac{d^3 k}{(2 \,\pi)^3} \,\frac{1}{\omega^2} \,
 \sum_{\lambda = 1}^2 \,\bm{k} \times \bm{\epsilon}_{\bm{k}, \lambda} \,
 e^{\,i \,\bm{k} \cdot \bm{x}} \,\bm{p} \cdot \bm{\epsilon}_{\bm{k}, \lambda}
 \,e^{\,- \,i \,\bm{k} \cdot \bm{q}} .
\end{eqnarray}
Then, with the use of the formula for the polarization sum, we get
\begin{equation}
 \sum_{\lambda = 1}^2 \,\bm{k} \times \bm{\epsilon}_{\bm{k}, \lambda} \,\,
 \bm{p} \cdot \bm{\epsilon}_{\bm{k}, \,\lambda} \ = \ 
 - \,\bm{p} \times \bm{k} ,
\end{equation}
we finally obtain
\begin{equation}
 \langle \tilde{0}_1 \,\vert \tilde{\bm{B}} (\bm{x}) \,\vert \tilde{0}_1 \rangle
 \ = \ - \,\frac{\bm{p}}{m} \times \nabla \,\frac{1}{4 \,\pi} \,
 \frac{1}{\vert \bm{x} - \bm{q} \vert} \ + \ O (e^2) ,
\end{equation}
which proves Eq.(95) in the text.


\begin{thebibliography}{}

\bibitem{ES1949}
W.~Ehrenberg and R. E.~Siday, 
{\it The refractive index in electron optics and the principles
of dynamics},
Proc. Phys. Soc. London {\bf B62}, 8 (1949).

\bibitem{AB1959}
Y.~Aharonov and D.~Bohm, 
{\it Significance of electromagnetic potentials in quantum theory},
Phys. Rev. {\bf 115}  (1959) 485.


\bibitem{Peshkin1981}
M.~Peshkin,
{\it The Aharonov-Bohm effect : Why it cannot be eliminated from
quantum mechanics},
Phys. Rep. {\bf 80} (1981) 375.

\bibitem{OP1985}
S.~Olariu and I. I.~Popescu,
{\it The quantum effects of electromagnetic fluxes},
Rev. Mod. Phys. {\bf 57} (1985) 339.

\bibitem{PT1989}
M.~Peshkin and A.~Tonomura,
{\it The Aharonov-Bohm effect},
Lecture Note in Physics {\bf 340}, (1989) 1.

\bibitem{WKZZ2018}
M.~Wakamatsu, Y.~Kitadono, L.~Zou, and P.-M. Zhang,
{\it The role of electron orbital angular momentum in the
Aharonov-Bohm effect revisited},
Ann. Phys. {\bf 397}  (2018) 259.


\bibitem{Feynman1963}
R. P.~Feynman, R. B.~Leighton, and M.~Sands,
{\it Feynman Lectures on Physics, Vol. II},
(Addison-Wesley Pub. Co., c1963-1965),
Ch.15.4.

\bibitem{Konopinski1978}
E. J.~Konopinski,
{What the electromagnetic potential describes?}, 
Am. J. Phys. {bf 46}  (1978) 499.

\bibitem{Semon-Taylor1996}
M. D.~Semon and J. R.~Taylor,
{\it Thoughts on the magnetic vector potential}, 
Am. J. Phys. {\bf 64}  (1996) 1361.



\bibitem{Tonomura1986}
A.~Tonomura, N.~Osakabe, T.~Matsuda, T.~Kawasaki,
and J.~Endo, 
{\it Evidence for Aharonov-Bohm effect with magnetic field completely
shielded from electron wave},
Phy. Rev. Lett. {\bf 56}  (1986) 792.

\bibitem{Osakabe1986}
N.~Osakabe, T.~Matsuda, T.~Kawasaki, J.~Endo, A.~Tonomura, S.~Yano,
and H.~Yamada,
{\it Experimental confirmation of Aharonov-Bohm effect using a
toroidal magnet field confined by a superconductor},
Phys. Rev. {\bf A34} (1986) 815.



\bibitem{Healey2022}
R.~Healey,
{\it Nonlocality and the Aharonov-Bohm Effect},
Philosophy of Science, {\bf 64} (1997) 18.

\bibitem{ACR2016}
Y.~Aharonov, E.~Cohen, and D.~Rhorlich,
{\it Nonlocality in the Aharonov-Bohm effect},
Phys. Rev. A {\bf 93} (2016) 42110.

\bibitem{Heras-Heras2022}
J. A.~Heras and R.~Heras,
{\it Topology, nonlocality and duality in classical electrodynamics},
Eur. Phys. J. Plus {\bf 137} (2022) 157.


\bibitem{Heras2022}
R.~Heras,
{\it The Aharonov-Bohm effect in a closed line}, 
Eur. Phys. J. Plus {\bf137}  (2022) 641.

\bibitem{EPR1935}
A.~Einstein, B.~Podolsky, and N.~Rosen,
{\it Can quantum-mechanical description of physical reality be
considered complete?},
Phys. Rev. {\bf 48}, 696 (1935).

\bibitem{Bell1964}
J.S.~Bell,
{\it On the Einstein Podolsky Rosen paradox},
Physics {\bf 1}, 195 (1964).


\bibitem{Adachi1992}
T.~Adachi, T.~Inagaki, M.~Ozaki, and K.~Sasabe,
{\it The Vector Potential Revisited},
Electrical Engineering in Japan, Vol. {\bf114}, No.6 (1993), 
Translated from Denki Gakkai Ronbunnshi, {\bf A112} (1992) 763.

\bibitem{Stewart2003}
A.M.~Stewart,
{\it Vector potential of the Coulomb gauge},
Eur. J. Phys. {\bf 24} (2003) 519.

\bibitem{Li2012}
J.-F.~Li, Y.~Jiang, W.-M. Sun, H.-S.~Zong, and F.~Wang,
{\it New application of decomposition of $U(1)$ gauge potential : 
Aharonov-Bohm effect and Anderson-Higgs mechanism},
Mod. Phys. Lett. {\bf B26}  (2012) 1250124.



\bibitem{Boyer1971}
T. H.~Boyer,
{\it Classical Electromagnetic Interaction of a Charged Particle with
a Constant-Current Solenoid}, 
Phys. Rev. {\bf 8} (1971) 1667.


\bibitem{Vaidman2012}
L.~Vaidman,
{\it Role of potentials in the Aharonov-Bohm effect},
Phys. Rev. A {\bf 86} (2012) 040101(R).

\bibitem{Kang2017}
K.~Kang,
{\it Proposal for locality test of the Aharonov-Bohm effect via
Andreev interferometer without a loop},
Phys. Soc. {\bf 71} (2017) 565.

\bibitem{Marletto-Vedral2020}
C.~Marletto and V.~Vedral,
{\it Aharonov-Bohm Phase is Locally Generated Like All Other
Quantum Phases},
Phys. Rev. Lett. {\bf 125} (2020) 040401.

\bibitem{Saldanha2021}
P. L.~Saldanha,
{\it Local Description of the Aharonov-Bohm effect with a Quantum
Electromagnetic Field},
Found. Phys {\bf 51}  (2021) 6.

\bibitem{Santos-Gonzalo1999}
E.~Santos and I.~Gonzalo,
{\it Microscopic theory of the Aharonov-Bohm effect}, 
Europhys. Lett. {\bf 45} (1999) 418.

\bibitem{LHK2022}
X.~Li, T. H.~Hansson, and W.~Ku,
{\it Gauge-independent description of the Aharonov-Bohm effect}, 
Phys. Rev. A {\bf 106} (2022) 032217.




\bibitem{Kobe1991}
D. H.~Kobe,
{Berry phase, Aharonov-Bohm effect and topology},
J. Phys. A : Math Gen. {\bf 24} (1991) 3551.

\bibitem{CLBNGK2019}
E.~Cohen, H.~Larocque, F.~Bouchard, F.~Nejadsattari, Y.~Gefen, and E.~Karimi,
{Geometric phase from Aharonov-Bohm to Pancharatnam-Berry and beyond}, 
Nature Review Physics {\bf 1} (2019) 437.



\bibitem{Heaviside1888}
O.~Heaviside,
{\it The electromagnetic effect of a moving charge},
The Electrician, {\bf 22}  (1888) 147.

\bibitem{Jefimenko1994}
O. D.~Jefimenko,
{\it Direct calculation of the electric and magnetic fields of an electric 
point charge moving with constant velocity}, 
Am. J. Phys. {\bf 62}  (1994) 79. 

\bibitem{RPS2006}
B.~Rothenstein, S.~Propescu, and G. J.~Spix,
{\it Relativistic derivations of the electric and magnetic fields generated
by an electric point charge moving with constant velocity},
arXiv:physics/0601028 (2006).







\bibitem{Cohen-Tannoudji1989}
C.~Cohen-Tannoudji, J.~Dupont-Roc, and G.~Grynberg,
{\it Photons and Atoms}, (Wiley, New York, 1989).


\bibitem{Feynman-Hibbs1965}
R. P.~Feynman and A. R.~Hibbs,
{\it Quantum Mechanics and Path Integral},
(McGraw-Hill Companies, Inc., NewYork, 1965).

\bibitem{Horvat2020}
S.~Horvat, P.A.~Gu\'erin, L,~Apadula, and F.~ Del Santo,
{\it Probing quantum coherence at a distance and Aharonov-Bohm
non-locality}, Phys. Rev. {A102} (2020) 062214.

\bibitem{Singleton-Vagenas2013}
D.~ Singleton and E.C.~Vagenas,
{\it The covariant, time-dependent Aharonov-Bohm effect}
Phys. Lett. {\bf B723} (2013) 241.

\bibitem{Shadowitz1975}
A.~Shadowitz,
{\it The Electromagnetic Field}, 
(Dover Publication, Inc., NewYork, 1975).


\bibitem{Babiker-Lowdon1984}
M.~Babiker and R.~Lowdon, {\it  Gauge invariance of the Aharonov-Bohm effect},
J. Phys. A: Math. Gen.  {\bf 17}, 2973 (1984).


\bibitem{WKZ2018}
M.~Wakamatsu, Y.~Kiadono, and P.-M.~Zhang,
{\it The issue of gauge choice in the Landau problem and the physics
of canonical and mechanical orbital angular momenta},
Ann. Phys. {\bf 392}  (2018) 287.

\bibitem{WH2022}
M.~Wakamatsu and A.~Hayashi,
{\it Physical symmetries and gauge choices in the Landau problem},
Eur. Phys. J. {\bf A58} (2022) 121.


\bibitem{BL1978}
P.~Bocchieri and A.~Loinger,
{\it Nonexistence of the Aharonov-Bohm effect},
Nuovo Cimento {\bf 47A}, 475 (1978).

\bibitem{Kreizschmar1965}
M.~Kreizschmar,
{\it Must Quantal Wave Functions be Single-Valued ?},
Zeitschrift F\"{u}r Physik {\bf 185}, 73 (1965).

\bibitem{BB1983}
M.~Bawin and A.~Burnel,
{\it Aharonov-Bohm effect and gauge invariance},
J. Phys. A: Math. Gen. {\bf 16} 2173 (1983).

\bibitem{Miyazawa2}
H.~Miyazawa and T.~Miyazawa, {\it The physical meaning of the Ehrenberg-Siday-Aharonov-Bohm effect}'' 
(in Japanese), http://www.miyazawa1.sakura.ne.jp/papers/esba.pdf.





\end{thebibliography}

\end{document}